\documentclass{emulateapj}
\usepackage{color}
\usepackage{multirow}
\font\cap=cmcsc10 
\slugcomment{Draft, \today}
\def\be{\begin{equation}}
\def\ee{\end{equation}}
\def\nms{\mathsurround=0pt}
\def\oversim#1#2{\lower 4pt\vbox{\baselineskip 0pt \lineskip 1pt
    \ialign{$\nms#1\hfil##\hfil$\crcr#2\crcr\sim\crcr}}}
\def\ga{\mathrel{\mathpalette\oversim>}}
\def\la{\mathrel{\mathpalette\oversim<}}
\def\arcdeg{{^{\circ}}}

\def\bh{M_{\bullet}}

\def\pc{{\rm ~pc}}
\def\kpc{\rm ~kpc}
\def\msun{M_{\odot}}
\def\ab{a_{\rm b}}
\def\abi{a_{\rm b,ini}}

\def\AU{{\rm AU}}
\def\kms{{\rm km\,s^{-1}}}
\def\tb{_{\rm tb}}

\def\HVS{_{\rm HVS}}

\def\ini{{\rm ini}}

\def\rp{r_{\rm p}}
\def\rpi{r_{\rm p,ini}}
\def\ac{a_{\rm cap}}
\def\ec{e_{\rm cap}}
\def\obs{^{\rm obs}}
\def\obsr{^{\rm obs,rf}}
\def\obsp{^{\rm obs,pm}}
\def\rf{_{\rm rf}}
\def\g{_{\rm g}}
\def\l{_{\rm l}}
\def\pyr{{\rm yr}^{-1}}
\def\maspyr{{\rm mas\,yr^{-1}}}
\def\vinf{v_{\infty}}
\def\vinfi{v_{\rm \infty,ini}}
\def\cap{{\rm cap}}
\def\ej{{\rm ej}}
\def\tot{{\rm tot}}
\def\HVS{{\rm HVS}}
\def\lt{{\rm lt}}
\def\td{{\rm td}}

\begin{document}
\shorttitle{Linking between GC S-stars and Galactic halo HVSs}
\shortauthors{Zhang, Lu \& Yu}

\title{THE GALACTIC CENTER S-STARS AND THE HYPERVELOCITY STARS IN THE
GALACTIC HALO: TWO FACES OF THE TIDAL BREAKUP OF STELLAR BINARIES
BY THE CENTRAL MASSIVE BLACK HOLE?}
\author{Fupeng Zhang$^1$, Youjun Lu$^1$, AND Qingjuan Yu$^{2}$}
\affil{$^1$~National Astronomical Observatories, Chinese Academy of Sciences,
Beijing, 100012, China \\
$^2$~Kavli Institute for Astronomy and Astrophysics, Peking University, 
Beijing, 100871, China 
}

\begin{abstract}

In this paper, we investigate the link between the hypervelocity stars
(HVSs) discovered in the Galactic halo and the Galactic center (GC)
S-stars, under the hypothesis that they are both the products of the
tidal breakup of the same population of stellar binaries by the
central massive black hole (MBH). By adopting several hypothetical
models for binaries to be injected into the vicinity of the MBH and
doing numerical simulations, we realize the tidal breakup processes of
the binaries and their follow-up dynamical evolution. We find that
many statistical properties of the detected HVSs and GC S-stars could
be reproduced under some binary injecting models, and their number
ratio can be reproduced if the stellar initial mass function is
top-heavy (e.g., with slope $\sim-1.6$). The total number of the
captured companions is $\sim50$ that have masses in the range
$\sim3$--$7\msun$ and semimajor axes $\la4000\AU$ and survive to the
present within their main-sequence lifetime.  The innermost one is
expected to have a semimajor axis $\sim300$--$1500\AU$ and a
pericenter distance $\sim10$--$200\AU$, with a significant probability
of being closer to the MBH than S2. Future detection of such a closer
star would offer an important test to general relativity. The majority
of the surviving ejected companions of the GC S-stars are expected to
be located at Galactocentric distances $\la20\kpc$, and have
heliocentric radial velocities $\sim-500$--$1500\kms$ and proper
motions up to $\sim5$--$20\maspyr$.  Future detection of these HVSs
may provide evidence for the tidal breakup formation mechanism of the
GC S-stars.

\end{abstract}


\section{Introduction}

More than a hundred young massive stars, mostly Wolf Rayet/O and B types, have
been identified within a distance of $\sim 0.5\pc$ from the massive black hole
(MBH) in the Galactic center \citep[GC;][]{Gillessenetal09,LuJ09,Bartko10}.
These young stars are empirically divided into two groups: (1) the
majority of the young stars at a distance $\sim 0.04$--$0.5\pc$ from the MBH are
located on coherent disk-like structures, i.e., the clockwise rotating stellar
(CWS) disk and the possible counterclockwise rotating stellar (CCWS) disk
\citep[e.g.,][]{LB03,Paumard06,LuJ09}; and (2) the young stars within a distance of
$0.04\pc$ from the MBH (denoted as GC S-stars), exclusively B-dwarfs, are
spatially isotropically distributed and their orbital eccentricities follow a
distribution of $f_e(e)\propto e^{2.6}$ \citep[e.g.,][]{Ghezetal08,
Gillessenetal09}. The existence of these young stars is quite puzzling as
star formation in the vicinity of an MBH is thought to be strongly suppressed
due to the tidal force from the MBH \citep[i.e., the paradox of youth;
see][]{Ghezetal08,Paumard06}. It is of great importance to address not only the
formation of these stars but also the origin of their kinematics, which should
encode fruitful information of the dynamical interplays between the central MBH
and its environment.

Young stars in the CWS (or CCWS) disk are probably formed in a previously
existing massive gaseous disk due to instabilities and fragmentation developed
in it \citep[e.g.,][]{Levin07,Nayakshin06,Alx08,Bonnell08}. Young binary stars
in the disk(s) may migrate or be scattered into the vicinity of the central MBH \citep[e.g.,][]{MLH09}
and then be tidally broken up \citep[e.g., ][]{Hills88,YT03}. One component of
a broken-up binary may be ejected out as a hypervelocity star (HVS) as
discovered in the Galactic halo \citep[e.g.,][]{Brown05,Edelmann05,Hirsch05},
and the other component may be captured onto a tighter orbit similar to that of
the GC S-stars as proposed by \citet{Gould03}.\footnote{
Some other scenarios were also proposed to explain the orbital configuration
of the GC S-stars, for example, dynamical interactions of these stars with an
intermediate-mass BH in the vicinity of the central MBH (see
\citealt{Merritt09,Gualandris09}) or migration of stellar binaries from the outer
stellar disk to the inner region and consequent supernova explosions (see
\citealt{BCL11}).} If HVSs were initially originated from a stellar
structure like the CWS disk, they may be spatially located close to the disk plane
\citep{Luetal10}.  The current observations do show such a spatial
correlation between the HVSs and the CWS disk, which suggests that majority of
the HVSs originate from the CWS disk \citep{Luetal10, Zhang10}. 

The HVSs
discovered in the Galactic halo and the GC S-stars in the vicinity of the central
MBH may naturally link to each other as they may both be the products of the tidal
breakup of stellar binaries in the vicinity of the central MBH \citep[e.g.,][]{GL06}.  Therefore, it
is interesting to simultaneously investigate the properties of the HVSs in the 
Galactic halo (or the GC S-stars) and their captured (or ejected) 
companions, and probability distribution of these properties.
Under the assumption that both the HVSs and GC S-stars are the products of
tidal breakup of stellar binaries, the working hypothesis in this paper, we
construct a number of Monte Carlo models to simulate the tidal breakup
processes of stellar binaries in the GC and check whether these models can
accommodate the current observations, and make further predictions on both the
companions of HVSs and that of GC S-stars for future observations.\footnote{In principle,
each HVS should have a companion left in the GC and each S-star should have a companion
ejected to the Galactic halo. However, these companions could have left the main sequence
because of the limited lifetime and cannot
be detected at the present time; and the captured companion of an HVS may even has been
tidally disrupted by the central MBH and does not exist now. Considering
of those cases, hereafter, the term ``companions'' may have a broad meaning in that
it includes the companions of previously existed HVSs or GC S-stars as well as those
detectable at the present time; and the companions of HVSs and GC S-stars may have
different numbers at the present time.}

This paper is organized as follows. In Section~\ref{sec:overview}, we overview
the tidal breakup processes of stellar binaries in the vicinity of an MBH and
the dynamical connection between the ejected and captured components.  Adopting
relatively realistic initial conditions, we perform a large number of
three-body experiments to realize the tidal breakup processes of stellar
binaries in Section~\ref{sec:num_simu}.  Assuming a constant injection rate of
stellar binaries into the vicinity of the central MBH and adopting the results
from the three-body experiments on the ejected and captured components, we use
the Monte Carlo simulations to produce both the HVSs and the GC S-stars. In
Section~\ref{sec:orb_ev}, we follow the orbital evolution of the captured stars
to the present time by adopting the autoregressive moving average (ARMA) model
\citep{MHL10}, in which both the non-resonant relaxation (NR) and the resonant
relaxation (RR) are included. The simulated GC S-stars appear to be compatible with
the observations of the GC S-stars. In Section~\ref{sec:HVSs}, we investigate the
effects of different binary injection models on the number ratio of the
simulated HVSs to GC S-stars. The number ratio given by observations can be
reproduced if the initial mass function (IMF) of the primary components of
stellar binaries is somewhat top-heavy. By calibrating the injection models
with observations, we estimate the number of the captured (or ejected unbound)
stars, as the companions of HVSs (or GC S-stars), that could be
detected in the future. We also estimate the probability to have less massive
stars captured on an orbit within that of S2 in Section~\ref{sec:innermost}. Conclusions are
given in Section~\ref{sec:conclusion}.

For clarity, some notations of the variables that are frequently used in this
paper are summarized in Table~\ref{tab:t1}. Given a physical variable $X$
(e.g., mass, velocity, semimajor axis, eccentricity), the distribution function of
$X$ is denoted by $f_X (X)$ so that $f_X(X)dX$ represents the number
of relevant objects with variable $X$ being in the range $X\rightarrow X+dX$.  

\begin{deluxetable*}{ll}
\tablewidth{18.5cm}
\tablecaption{Notation of Some Symbols}
\tablehead{ \colhead{Symbol} & \colhead{Description} }
\startdata
\tabletypesize{\tiny}
$\bh$                   & Mass of the central MBH                                        \\
$m_{\rm p}$             & Mass of the primary component of an injecting stellar binary   \\
$m_{\rm s}$             & Mass of the secondary component of an injecting stellar binary \\
$m$                     & Total mass of an injecting stellar binary, i.e., $m_{\rm p}+m_{\rm s}$     \\
$R$                     & $m_{\rm s}/m_{\rm p}$                                          \\
$\abi$                  & Initial semimajor axis of an injecting stellar binary         \\
$\rpi$                  & Initial pericenter distance of the mass center of the injecting stellar binary to the MBH \\
$a_{\rm b-\bullet,ini}$ & Initial semimajor axis of the orbit of an injecting stellar binary rotating around a central MBH \\
$\vinfi$                & Initial velocity of the injecting stellar binary at infinity if the binary is on a hyperbolic orbit \\
$E_\ini $               & Initial energy of the stellar binary   \\
$r\tb$                  & Tidal radius for the stellar binary \\
$D$                     & Orbital penetration parameter of the injecting stellar binary ($\equiv 100\rpi/r\tb$)\\
$\alpha$                & Exponent of the power-law distribution of $\abi$      \\                                                        
$\beta$                 & Exponent of the power-law distribution of $\rpi$      \\                                                        
$\gamma$                & Exponent of the power-law distribution of $m_{\rm p}$ \\ 
                        &   \\ \hline
$m\g $                  & Mass of the component that gains energy during the tidal breakup of a stellar binary\\
$m\l $                  & Mass of the component that loses energy during the tidal breakup of a stellar binary\\
$q$                     & $m\l/m\g$                                                   \\
$\delta E$              & Exchange energy between the two components during the tidal breakup of a stellar binary \\ 
                        &   \\ \hline
$m_{\rm ej}$            & Mass of the ejected star after the tidal breakup of a stellar binary     \\   
$m_{\cap}$              & Mass of the captured star after the tidal breakup of a stellar binary     \\ 
$\vinf$                 & Velocity of the ejected component at infinity \\
$a_{\cap} $             & Orbital semimajor axis of the captured component \\
$a_{\rm cap,0}$         & Orbital semimajor axis of the captured component if the injecting binary is initially on a parabolic orbit \\   
$e_{\cap} $             & Orbital eccentricity of the captured component \\
$N_{\HVS}^{\tot}$       & Simulated total number of the ejected stars given a mass range\\  
$N_{\cap}^{\tot}$       & Simulated total number of the captured stars given a mass range\\
$N_{\HVS}\obs$          & Simulated number of the detectable HVSs at the present time for given selection criteria\\
$N_{\cap}\obs$          & Simulated number of the detectable captured stars at the present time for given selection criteria \\
$F_{\HVS}^{\lt}$        & Simulated fraction of the ejected stars that survive to the present time on the main sequence  \\
$F_{\cap}^{\lt}$        & Simulated fraction of the captured stars that survive to the present time on the main sequence  \\
$F_{\cap}^{\td}$        & Simulated fraction of the captured stars that have already been tidally disrupted until the present time \\
$F_{\cap}\obs$          & Simulated fraction of the captured stars that can be detected at the present time for given selection criteria\\
\enddata
\label{tab:t1} 
\end{deluxetable*}

\section{Overview: tidal breakup of stellar binaries in the vicinity of an MBH}
\label{sec:overview}

A stellar binary may be broken up if it approaches an MBH within a distance
of $r\tb=\ab(3M_\bullet/m)^{1/3}$, where $M_\bullet$ is the mass of the MBH,
$\ab$ is the semimajor axis of the binary, $m=m\g+m\l$ is the total mass of the
binary, and $m\g$ and $m\l$ are the masses of the two components of the binary,
respectively. During the breakup, one component of the binary, denoted as
$m\g$ here, gains energy, and the other component $m\l$ loses energy. For an
injecting stellar binary that is initially on a parabolic orbit relative to the
MBH, the velocity of the binary mass center at its periapsis to the MBH ($\sim
r\tb$) is $v\tb\sim (GM_\bullet/r\tb)^{1/2}$. The component $m\g$ receives
a velocity change on the order of $\delta v\g \sim (m\l/m)\sqrt{Gm/ \ab}$ if the
eccentricity of the stellar binary is $0$, and it gains energy $\delta E\sim m\g
v\tb\delta v\g$. The other component $m\l$ loses the same amount of energy
$\delta E$. If $\delta E$ is sufficiently large, the component $m\g$ may
manifest itself as an HVS with velocity at infinity $\vinf\sim \sqrt{2
\delta E/m\g}$ if ignoring the deceleration due to the Galactic gravitational potential. The
root mean square (rms) of $\vinf$ is approximately
\begin{eqnarray}
\left<v^2_{\infty}\right>^{1/2} & \sim & v_{\infty,0}
\left(\frac{0.1\AU}{\ab}\right)^{1/2}
\left(\frac{m}{6\msun}\right)^{1/3}\nonumber\\
&\times &\left(\frac{2m\l}{m}\right)^{1/2}
\left(\frac{M_\bullet}{4\times 10^6\msun}\right)^{1/6} g(D),
\label{eq:vhvs}
\end{eqnarray}
where $ v_{\infty,0}=2596\kms$ and $g(D)$ is given by
\citet[][]{Bromley06} for injecting binaries on hyperbolic orbits with initial
velocities at infinity of $250\kms$, i.e.,
\begin{eqnarray}
g(D)& = & 0.774+0.0245D-8.99\times 10^{-4}D^2 \nonumber \\
& & +1.32\times 10^{-5}D^3-8.82\times10^{-8}D^4 \nonumber \\
& & +2.15\times 10^{-10}D^5,
\label{eq:fd1}
\end{eqnarray}
where the penetration parameter $D\equiv 100\rpi/r\tb$ characterizes the
minimum distance where the binary approaches the MBH, and $\rpi$ is the initial
pericenter distance of the binary.
The rms velocity $\left<v^2_{\infty}\right>^{1/2}$ apparently depends on the
semimajor axis, the total mass and the mass ratio of the stellar binary, and the
penetration parameter $D$.

The exact value of $\vinf$ of the ejected component for any given stellar
binary also depends on the relative orientation of the stellar binary orbital
plane to the orbital plane of the binary rotating around the MBH and the
orbital phases of the two components at the time of its breakup. This dependence
introduces a scatter of $\vinf$ around the value $\left< v^2_{\infty}
\right>^{1/2}$ given by Equation (\ref{eq:vhvs}), as the orbital orientations of
the injecting stellar binaries are probably random and the orbital phases of
the two components are not fixed at the breakup time.  Numerical simulations
have shown that this scatter is approximately Gaussian with a dispersion of
$\sigma_{\vinf}\sim 0.2 \left<v^2_{\infty} \right>^{1/2}$
\citep{Bromley06,Zhang10}, where the binary orbital orientations are assumed to
be randomly distributed. The symmetry of the orbital phases of the two binary
components (always at the opposite side to the mass center of the binary)
ensures the same probability of receiving energy for each star, which leads to
the same ejection probability for both components if the injecting binaries are
initially on parabolic orbits \citep{Sari10,Sari12}.

Stellar binaries on orbits bound to the MBH may experience multiple close
encounters with the MBH and the binary semimajor axes and eccentricities may be
cumulatively excited to larger values until finally being broken up
\citep{Zhang10}. The distribution of $\vinf$ for the ejected stars,
produced during the first encounters of the binaries with the MBH, follows a
fitting formula similar to Equation (\ref{eq:fd1}) over $D\sim 20$--$150$, i.e.,
$g(D) \propto 1-(D/256)^2$, as the initial bounding energy of the
injecting stellar binaries is still significant~\citep[for details,
see][]{Zhang10}.  For multiple encounters, the energy exchange $\delta E$
between the two components is determined by the properties of the stellar
binaries at the final revolutions. Our simulations show that
$\left<v^2_{\infty}\right>^{1/2}$ of those ejected components for stellar
binaries broken up within $1000$ revolutions around the MBH still follows
Equation (\ref{eq:vhvs}), but $g(D)$ is now best fitted by
\begin{eqnarray}
g(D)& = &0.912-2.41\times10^{-4} D -4.49\times10^{-5}D^2\nonumber \\
& &+2.68\times10^{-7} D^3-4.42\times10^{-10} D^4, 
\label{eq:fd2}
\end{eqnarray}
for $D<300$. 

For stellar binaries on bound orbits, the light component has a larger
probability to escape away from the MBH because the specific energy it could
gain is generally larger than that of the heavy component in a counterpart case
(see Equation~\ref{eq:vhvs}). However, the difference in the ejection probability
for the two components of the stellar binaries is significant only when the
mass ratio of the massive ones to the light ones $\ga 5$ and $2\delta E/m\g$ is
close to its initial bounding energy $GM_{\bullet}/2a_{\rm b-\bullet,\ini}$, where
$a_{\rm b-\bullet,\ini}$ denotes the initial semimajor axis of the binary system
composed of a stellar binary and the MBH \citep[see also][]{Antonini11, Sari12}.

The component $m\l$ of a broken-up stellar binary loses energy by an amount of
$\delta E$ and it is captured onto a tighter orbit with semimajor axis $\ac$.
According to the energy conservation law, we roughly have
\be 
\frac{1}{2}m\g v_{\infty}^2-\frac{Gm\l\bh}{2\ac} \simeq E_{\ini},
\label{eq:dk_b} 
\ee 
where $E_{\ini}$ is the initial energy of the stellar binary,
and it is
$\sim \frac{1}{2}m v_{\infty,\ini}^2$ 
if the binary is initially
on a hyperbolic orbit, or $\sim -Gm\bh/(2a_{\rm b-\bullet,\ini})$ if 
on a bound orbit. The initial internal mechanical energy
of the stellar binary $-\frac{Gm_lm_g}{2a_{\rm b,ini}}$ is ignored in Equation (\ref{eq:dk_b}).
We now have the general form for $\vinf$ as
$\vinf \sim \sqrt{2(\delta E+\frac{m\g}{m} E_{\ini})/m\g}$.
If $|E_{\ini}|\ll \delta E$, the semimajor axis of the captured star is 
\begin{eqnarray}
\ac& \simeq  & a_{\rm cap,0}  =  q\frac{G\bh}{v^2_\infty} \nonumber \\
            & = & 3500q \AU 
                  \left(\frac{M_{\bullet}}{4\times 10^6\msun}\right)
                  \left(\frac{1000\kms}{\vinf}\right)^2, 
\label{eq:acap0}
\end{eqnarray}
where $q\equiv m\l/m\g$.
For the cases considered in this paper, the injecting stellar binaries are
either initially on hyperbolic orbits (but close to parabolic ones) or from
stellar structures like the CWS disk, and thus $\ac\sim a_{\rm cap,0}$ as
approximately $|E_{\ini}|\ll \delta E$. Equation (\ref{eq:acap0}) shows the
connection between the properties of the captured stars left in the GC and that
of their ejected companions in the Galactic bulge and halo. For those HVSs
discovered in the Galactic halo with $\vinf \sim 700$-$1000\kms$,\footnote{
The estimated $\vinf$ for those detected HVSs depends on the Galactic
potential model adopted, the values here are obtained from the Galactic
potential model given by \citet{Xue08}.} their companions left in the GC may
be initially on orbits with semimajor axis in the range of $\sim
3500$-$7000\AU$ as stellar binaries with extreme mass ratios are rare. For the
innermost S-star, i.e., the S2, of which the semimajor axis is $\sim 1000\AU$, its
companion ejected out should have $\vinf \sim 1900\kms$ if $q \sim1$,
and $\sim 600\kms$ if $q \sim 0.1$, respectively. For stellar binaries
initially tightly bound to the MBH, we may have $\delta E + \frac{m\g}{m}
E_{\ini}\la 0$, the component gaining energy either remains bound to the
MBH or is ejected out with low velocities. 

The distribution of HVS properties is directly connected to the distribution of
S-star properties (note that here we do not mean that an observed HVS in
the Galactic halo is directly associated with an observed GC S-star as
the products of the tidal breakup of the same binary star).
According to Equation (\ref{eq:acap0}), the distribution of the semimajor axis
of the captured stars $f_{\ac}$ is related to the distribution of the
velocity of HVSs at infinity $f_{\vinf}$ if $m\g\sim m\l$, i.e., 
\be 
f_{\vinf}(\vinf)\propto \left.
\ac^{3/2}f_{\ac}(\ac)\right|_{\ac=\frac{G\bh}{v_\infty^2}},
\label{eq:va}
\ee
which suggests that any one of the two distributions above can be inferred
from the other one. 

The velocity distribution of the ejected stars are mainly determined by the 
initial sets on the distributions of $\abi$ and $\rpi$ since
\be 
\left<v^2_{\infty}\right>^{1/2}\propto \abi^{-1/2}g(D),
\ee 
where $g(D)$ denotes the dependence of the rms velocity
$\left<v^2_{\infty} \right>^{1/2}$ on the penetration parameter $D$, as shown
in Equations (\ref{eq:fd1}) and (\ref{eq:fd2}) for the cases of injecting binaries
initially on hyperbolic orbits but close to parabolic ones and bound orbits
like the stars in the CWS disk, respectively. The fitting forms of $g(D)$ obtained 
from numerical experiments (Equations (\ref{eq:fd1}) and (\ref{eq:fd2}))
are decreasing functions in the range of $20<D<150$ or $20<D<300$, and thus
$g(D)$ may be approximated as a monotonically decreasing function. We assume
that the initial distribution of $\abi$ and $\rpi$ are $f_{\ab}(\abi) \propto
\abi^\alpha$ and $f_{\rp}(\rpi) \propto \rpi^\beta$, respectively; 
and the probability of a stellar binary with semimajor axis $\abi$
broken up by the central MBH at a penetration distance $D$ is only a function of $D$, i.e., $f_D(D)$
\citep[see][]{Bromley06}. If $q\sim 1$ and ignoring the scatter of $\vinf$ 
around $\left<v_{\infty}^2\right>^{1/2}$ (i.e., $\vinf\sim \left<v^2_{\infty}\right>^{1/2}$), 
the velocity distribution of the ejected components
can be obtained as
\begin{eqnarray}
 f_{\vinf}(\vinf) & \propto &  \frac{\partial}{\partial v_\infty}\int\int f_{\ab}(\abi)f_{\rp}(\rpi) \times \nonumber \\
                  &         &  f_D(D) d\abi d\rpi \nonumber \\
                  & \propto & \frac{\partial}{\partial v_\infty} \int a_{\rm b,ini}^{\alpha+\beta+1} d\abi \int 
                  D^{\beta} f_D(D) dD \nonumber \\
                  & \propto &v_\infty^{-2\alpha-2\beta-5},
\label{eq:fV}
\end{eqnarray}
and this relation is valid only if $g(D)$ is a monotonically decreasing function
and it is independent of the detailed form of $f_D(D)$. Similarly, we also have
\be 
f_{\ac}(\ac) \propto \ac^{\alpha+\beta+1},
\label{eq:fac} 
\ee 
which is consistent with the simple relation given by Equation (\ref{eq:va}).
The estimated slope of $f_{\vinf}(\vinf)$ (or $f_{\ac}(\ac)$) above is
not affected by taking account of the Gaussian-like scatter of $\vinf$ around
$\left<v^2_{\infty}\right>^{1/2}$ as the distribution is a power law. If considering of
the various mass ratios among the injecting binaries (see Equations (\ref{eq:vhvs})
and (\ref{eq:acap0})), however, the resulted slope of $f_{\vinf}(\vinf)$ 
may be somewhat flatter than the simple estimates above.
Note also that a larger $\beta$ may correspond to a slower migration/diffusion
of stellar binaries into the low angular momentum orbits or the vicinity of the central MBH, 
and lead to fewer
HVSs at the high-velocity end and fewer captured stars in smaller distances to
the MBH.

The periapsis of a captured star $m\l$ is roughly $\sim r\tb$ and 
the characteristic eccentricity of the captured star is
\be 
\bar{e}_{\cap}\sim 1-\frac{r\tb}{\ac}\simeq 1- \frac{2.8}{q^{1/3}(1+q)^{2/3}} 
\left(\frac{m\l}{M_{\bullet}} \right)^{1/3}.
\label{eq:ecc} 
\ee 
the $\bar{e}_{\cap}$ depends on $q$ and the mass ratio of the captured star to 
the MBH. Considering of the Gaussian-like scatter in $\vinf$ and
correspondingly the scatter in $\ac$, the probability that the breakup
of a stellar binary with given semimajor axis and mass of each component 
results in a captured star with eccentricity $<\ec$ is roughly
\begin{eqnarray}
P(<\ec)=\frac{1}{2}{\rm erfc}\left[\frac{\sqrt{(1-\ec)/ 
(1-\bar{e}_{\cap})}-1}{\sqrt{2}\sigma_{\vinf}/ \langle v^2_{\infty}\rangle^{1/2}}\right], 
\label{eq:ecprob}
\end{eqnarray}
where $\bar{e}_{\cap}$ is given by Equation (\ref{eq:ecc}) and
$\sigma_{\vinf}/ \langle v^2_{\infty}\rangle^{1/2}\simeq 0.2$.
To capture an S2-like
star (i.e., $\ec\simeq 0.887$ and $m\l\sim 15\msun$; see \citealt{Ghezetal08};
\citealt{Gillessenetal09}) directly through the tidal breakup of stellar
binaries, it is necessary to have $m\g\gg m\l \sim 15 \msun$ ($q\ll 1$) and the
probability is $\sim 0.25$ if $q=0.1$ according to Equation (\ref{eq:ecprob})
\citep[see][]{Gould03}.  The probability to capture stars onto orbits with
$\ec<0.8$ is only $\sim 10^{-17}$ if $m\g=m\l=15\msun$; and $\sim6 \times
10^{-3}$ even if $m\g=10m\l=150\msun$ (see also \citealt{Gould03}). Since the
number of GC S-stars is only on the order of a few tens, it is difficult
to produce all the nine observed GC S-stars with eccentricities $<0.8$
\citep{Gillessenetal09} directly by the tidal breakup of stellar binaries.  In
addition, the captured stars may initially remain on a disk plane if their
progenitor binaries are originated from disk-like stellar structure(s) as
suggested by \citet{Luetal10}, which is different from the isotropic
distribution of the GC S-stars. Therefore, additional physical mechanism is
required to further make the captured stars evolve to orbits with lower
eccentricities and spatially isotropically distributed if the GC S-stars are
really originated from the tidal breakup of binary stars.

The processes, initially proposed by \citet{RT96}, may
cause the captured stars dynamically evolving to their present orbits as
discussed by a number of authors \citep{Levin07,HA06,KT11}. In
Section~\ref{sec:orb_ev}, we will take into account the relaxation
processes, including RR, to approximately follow the dynamical evolution of
each ``GC S-star'' after its capture due to the tidal breakup of stellar binaries;
we then check whether the eccentricity and spatial distributions of those
surviving ``GC S-stars'' are compatible with current observations. Note here we
use the quotes around the term GC S-stars to represent all of those captured stars
with mass in the range of $\sim 7$-$15\msun$;
while the simulated GC S-stars (without quotes) represent those with
mass $\sim 7$-$15\msun$ surviving to the present time, which presumably correspond to the observed ones (see Section~\ref{sec:orb_ev}).

\section{Monte Carlo Simulations}\label{sec:num_simu}

In this section, we first adopt Monte Carlo simulations to realize the tidal
breakup processes and generate HVSs and ``GC S-stars'', and then we investigate in
detail the connection between the simulated HVSs and ``GC S-stars''. We use the
code DORPI5 based on the explicit fifth (fourth)-order Runge--Kutta method
\citep{DP80,Hairer93} to calculate the three-body interactions between a stellar
binary and the central MBH.  For details of the numerical calculations, see
\citet{Zhang10}. The successive dynamical evolution of the captured stars in
the GC and the kinematic motion of the produced HVSs in the Galactic potential
will be discussed in Sections~\ref{sec:orb_ev} and \ref{sec:HVSs}, respectively.

\subsection{Initial Settings}\label{subsec: init_set}

The mass of the central MBH is set to be $4\times10^6\msun$ throughout
the numerical calculations in this paper \citep{Ghezetal08,Gillessenetal09}.

For the injecting stellar binaries, the initial conditions are set as 
follows: 

\begin{itemize}

\item The distribution of the semimajor axes $a_{\rm b,\ini}$ follows
the \"{O}pik law, i.e., $\alpha=-1$~\citep[e.g.,][]{KF07}.  

\item The mass distribution of the primary stars $m_{\rm p}$ follows a
power law function, $f_{m_{\rm p}}(m_{\rm p}) \propto m_{\rm p}^{\gamma}$. The distribution of
the secondary star ($m_{\rm s}$) or the mass ratio $R\equiv m_{\rm s}/m_{\rm p}$ can
be described by two populations: (1) a twin population, i.e., about 40\% of binary
stars have $R\sim 1$, and (2) the rest binaries, which follow a distribution of
$f_R(R)\sim {\rm constant}$~\citep{KF07,Kiminki08,Kiminki09}. 

\item The initial eccentricity of the injecting binary is assumed to be
$e_{\ini}=0$, as adopted in previous works (e.g., \citealt{Bromley06};
\citealt{Antonini10}).\footnote{Alternatively assuming the initial
eccentricities $\sim 0.3$, the velocities of the resulted HVSs from the four
models are roughly smaller than those obtained for $e_{\ini}\sim 0$ by 
$\la 10\%$.}

\item The orientation of the inner binary orbital plane is chosen to 
be uniformly distributed in $\cos\phi$ for $\phi\in (0,\pi)$.

\end{itemize}

For the orbits of the injecting stellar binaries relative to the
central MBH, the initial conditions are set as follows: 

\begin{itemize}

\item The stellar binaries are assumed to be initially injected from either
disk-like stellar structures (similar to the CWS disk) or infinity. If they
were from structures like the CWS disk, the semimajor axes $a_{\rm b-\bullet,\ini}$
follows a power-law distribution proportional to $a_{\rm b-\bullet,ini}^{-2.3}$ in the
range of $\sim 0.04$-$0.5\pc$ according to current observations on the CWS disk
\citep{LuJ09,Bartko09}.  If they were from infinity, i.e., unbound to the MBH,
their initial velocities at infinity are set to $\vinfi=250\kms$.  

\item If the injecting binaries were from disk-like stellar structures, the
orientations of their orbits relative to the MBH are assumed to satisfy a
Gaussian distribution around the central planes of the stellar disks with a
standard deviation of $12\arcdeg$ \citep[cf.][]{LuJ09,Bartko09}.  The planes
of the host disks are assumed to be the same as the two planes that best fit
the observations, i.e., $(l,b) = (311\arcdeg, -14\arcdeg)$ and $(176\arcdeg,
-53\arcdeg)$, respectively, and these two planes are consistent with the CWS
disk plane and the plane of the northern arm (Narm) of the mini-spiral in the GC (or the outer
warped part of the CWS disk;
\citealt{Luetal10, Zhang10}).\footnote{Note that \citet{Brown12} recently reported five new unbound HVSs discovered in the Galactic halo and
re-analyzed the HVSs previously discovered. According to this new study, there are $17$ unbound HVSs in the northern sky and
they are still consistent
with being located on two planes revealed by \citet{Luetal10}. That is, one of the disk planes is consistent with the CWS disk plane,
while the other disk plane is more consistent with the warped outer part of the CWS disk and slightly
deviates from the Narm plane.
In this paper, we do not distinguish the Narm plane from the warped outer part of the CWS disk.} The injection rates from these two disks are assumed
to be the same.

\item The periapsis that the injecting stellar binaries approach the MBH is
simply assumed to follow a power law distribution,
$f_{\rp}(\rpi)\propto\rpi^{\beta}$ and $\beta>0$. A larger value of $\beta$
corresponds to a smaller fraction of the injecting stellar binaries that could
approach the immediate vicinity of the central MBH. It is still not clear
which mechanism is responsible for the migration (or diffusion) of stellar
binaries into the vicinity of the central MBH, although the secular instability
developed in a stellar disk is proposed to be a viable one \citep{MLH09}.
Instead of incorporating the detailed migration/diffusion
process of the stellar binaries in the Monte Carlo simulations below, we choose
to parameterize the migration/diffusion process qualitatively by different values
of $\beta$ and a larger $\beta$ corresponds to a slower migration/diffusion
process.

\end{itemize}

\begin{deluxetable}{lcccc}
\tablecaption{Different Injection Models for Tidal Breakup of Binaries}
\tablehead{ \colhead{Model} & \colhead{$\gamma$} & \colhead{$\beta$}
& \colhead{$a_{\rm b-\bullet,ini}(\pc)$} & \colhead{$\vinfi(\kms)$\tablenotemark{a}}} 
\startdata
Unbd-MS0 & -2.7  & 0 & $\cdots$ & 250      \\
Disk-MS0    & -2.7  & 0 & 0.04-0.5 & $\cdots$ \\
Disk-TH0    & -0.45 & 0 & 0.04-0.5 & $\cdots$ \\
Disk-TH2    & -0.45 & 2 & 0.04-0.5 & $\cdots$
\enddata
\tablenotetext{a}{For the Unbd-MS0 model, the injecting stellar binaries
have initial velocities of $250\kms$ at infinity.} 
\label{tab:t2}
\end{deluxetable}

In this section, we perform Monte Carlo simulations by adopting four sets of
initial conditions (as listed in Table~\ref{tab:t2}). In the first
model, the stellar binaries are assumed to be injected from infinity with
initial velocity of $\vinfi =250\kms$. For the primary components of
the injecting binaries, we adopt the Miller Scalo IMF (e.g., \citealt{Kroupa}).
This model is denoted as ``Unbd-MS0''. For the other three models, the
stellar binaries are assumed to be originated from stellar structures like the
CWS disk, and the IMF of the primary components is either set to be the
Miller Scalo IMF or a top-heavy IMF with a slope of $\gamma=-0.45$ as suggested
by recent observations of the disk stars \citep[see][]{Bartko10}. The slope of
the initial distribution of the pericenter distance $\beta$ is set to be either
$0$ or $2$. These models are denoted as ``Disk-MS0'', ``Disk-TH0'', and
``Disk-TH2'', respectively. The total number of three-body experiments is $10^5$
for each model with the initial settings described above. By comparing the
results obtained from those different models, one may be able to distinguish the
effects of different settings on the IMF and the injection of stellar binaries.

If not specified, those ejected or captured stars with mass in the
range of $\sim 3$-$15\msun$ are recorded, thus both the HVSs with mass
$\sim 3$-$4\msun$ and the captured stars with mass $\sim 7$-$15\msun$,
corresponding to the currently detected ones, can be taken into
account simultaneously. The ejected or captured stars with mass in the
range of $\sim 4$-$7\msun$ are also considered for completeness.  For
other ejected or captured stars with mass out of the range of
$3$-$15\msun$, they may be either too faint to be detected or too
massive with too short lifetime and thus with too small probability to
survive. 

\subsection{Numerical Results}\label{subsec:results}

\begin{figure*}
\centering
\includegraphics[scale=0.5]{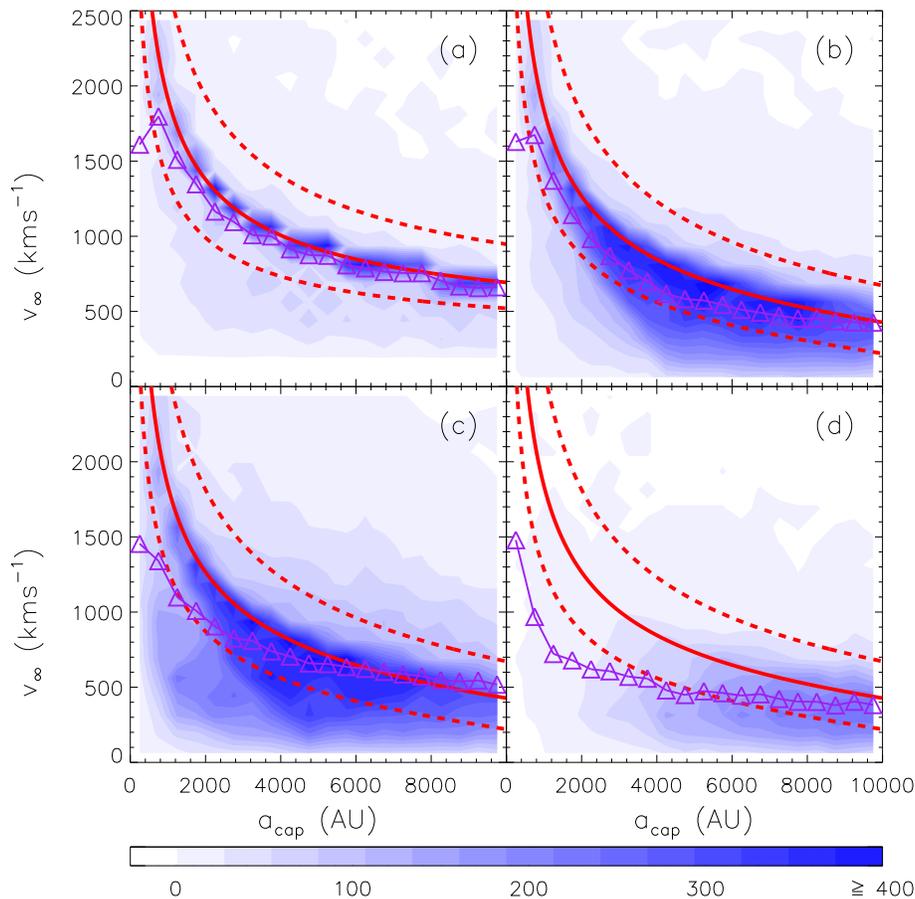} 
\caption{
Number distributions of broken-up stellar binaries in the $\vinf$-$\ac$
plane, where $\vinf$ is the velocity at infinity of the ejected component
and $\ac$ is the semimajor axes of its captured companion. Panels
(a)--(d) show
results obtained from the Unbd-MS0, Disk-MS0, Disk-TH0, and Disk-TH2 models,
respectively. The total number of the three-body experiments is $10^5$ for each
model. The solid red line in each panel shows the estimation according
to Equation (\ref{eq:dk_b}) for $m\g=m\l$ by assuming $\vinfi =
250\kms$ in panel (a), and $a_{\rm b-\bullet,\ini}=0.2\pc$ in panels
(b)--(d), respectively. The dashed
red lines above or below the solid lines represent the estimations for
$m\g/m\l=1/2$ and $2$, respectively. The magenta lines with triangles indicate
the rms of $\vinf$ for each bin of $\ac$.  The number of stars is counted
in each of the $\ac$ and $\vinf$ bins (totally $25\times25$ bins with bin size $400\AU\times 100\kms$) and represented by the color brightness scales
shown in the label.
} 
\label{fig:f1}
\end{figure*}

\begin{figure*}
\centering
\includegraphics[scale=0.5]{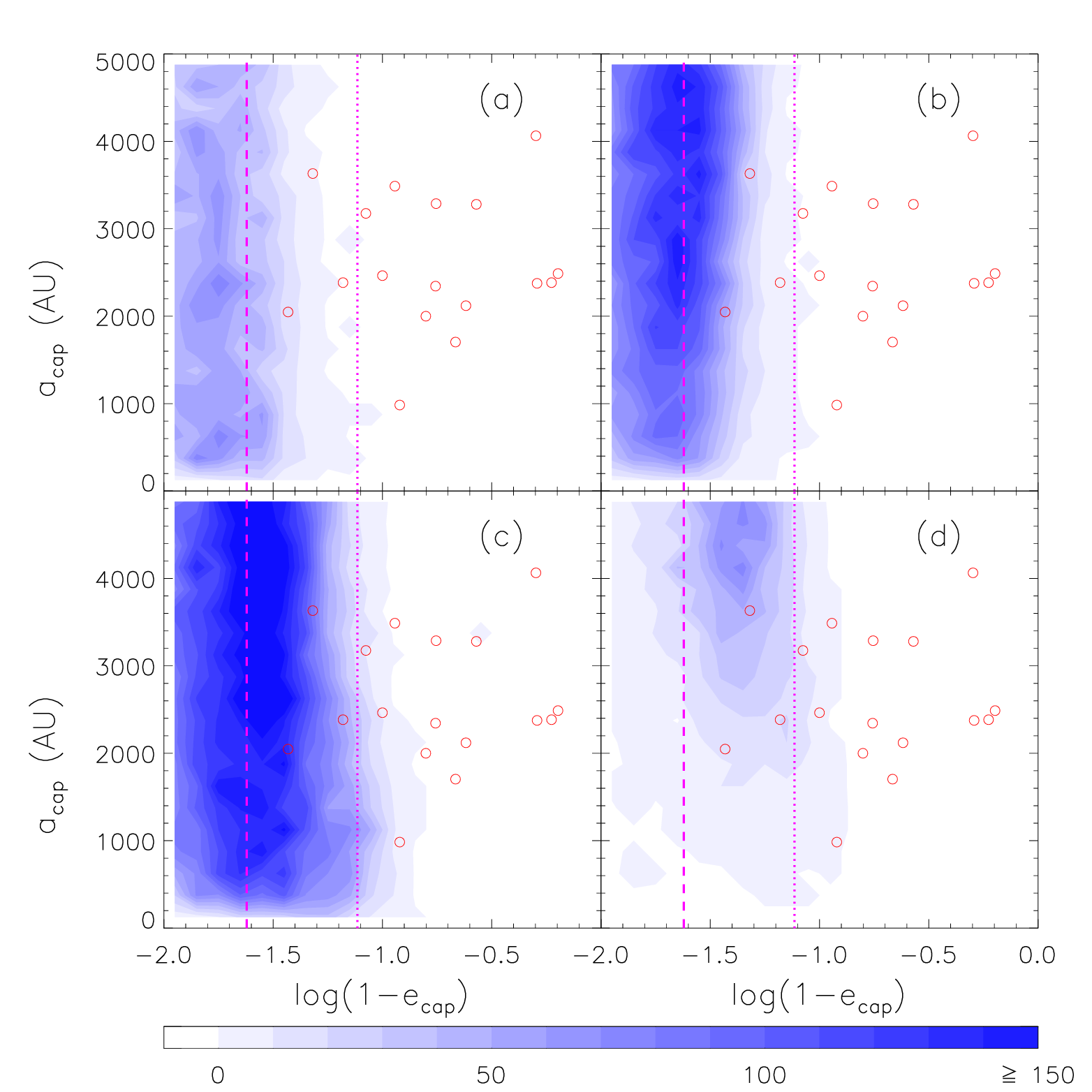}
\caption{
Number distributions of the captured stars in the $\ac$ vs. $\log(1-\ec)$ plane,
where $\ac$ and $\ec$ are the semimajor axes and eccentricities of the captured
stars achieved right after their capture, respectively.  Panels (a)--(d) show
the results from the Unbd-MS0, Disk-MS0, Disk-TH0, and Disk-TH2 model,
respectively.  The red open circles represent the observed GC S-stars with $\ac$ smaller than
$4000\AU$ at the present time \citep{Gillessenetal09}; and the magenta dashed
and dotted lines are for the mean eccentricities given by Equation
(\ref{eq:ecc}) for $(m\l, q)=(10\msun, 1)$ and $(10\msun, 1/10)$, respectively.
The number of stars is counted in each of the $\ac$ and $\log(1-\ec)$ bins
(totally $25\times25$ bins with bin size $200\AU\times 0.08$) and represented by the color brightness scales
shown in the label.
} 
\label{fig:f2} 
\end{figure*}

Figure~\ref{fig:f1} shows the distribution of the tidally broken-up stellar
binaries in the $\vinf$-$\ac$ plane, where $\vinf$ is the velocity at
infinity of the ejected component and $\ac$ is the semimajor axis of the
captured component. As shown in panel (a), the majority of the simulated
$\vinf$-$\ac$ pairs obtained from the Unbd-MS0 model are close to the one
estimated from Equation (\ref{eq:acap0}) by setting $m\l/m\g=1$ (solid line). The
main reasons for this are: (1) the majority ($70\%$) of the injecting stellar
binaries have mass ratios $q=m\l/m\g$ in the range of (1/2, 2) under the assumption of
two populations set for the stellar binaries; and (2) all the injecting binaries
have the same but negligible initial energy $E_{\ini}$. A small number of
$\vinf$-$\ac$ pairs, which apparently deviate significantly away from the
solid line (for $q=1$ obtained from Equation \ref{eq:acap0}), are due to the
breakup of the binaries with $q$ substantially larger or smaller than $1$ (below or
above the solid line). For the other three models, the simulation results
do not deviate far away from the simple predictions by Equation (\ref{eq:dk_b}) (for
$q=1$), except that fewer ejected stars at the high-velocity end are produced in the
Disk-TH2 model than in the other models  simply because not many stellar
binaries can closely approach the MBH. The scatters of $\vinf$ around
that predicted by Equation (\ref{eq:dk_b}) in panels (b)-(d) are more significant
compared with that in panel (a), which is caused by one or the combination of the
effects as follows: (1) a distribution of the negative initial energy of the
injecting stellar binaries originated from stellar structure like the CWS disk
(panels (b)-(d)); (2) relatively more progenitor binaries have $q$ substantially
larger or smaller than $1$ in the cases with a top-heavy IMF (panels
(c) and
(d)); and (3) fewer stellar binaries approach the immediate vicinity of
the central MBH in the case of a large $\beta$ (panel (d)).

As seen from Figure~\ref{fig:f1}, if the observed GC S-stars, with
semimajor axes $\sim 1000$-$4000\AU$, are produced
by the tidal breakup of stellar binaries, their ejected companions are expected
to have $\vinf \sim 1000$-$1600\kms$ in the Unbd-MS0 model and $\sim
200$-$1500\kms$ in the other models. The Disk-TH2
model produces fewer ejected stars with $\vinf$ substantially larger than
$1000\kms$ compared with other models. The captured companions of the detected
HVSs in the Galactic halo are more likely to have $\ac\sim 3000$-$8000\AU$ in
the Unbd-MS0 model, which is consistent with the simple estimation by
Equation (\ref{eq:acap0}), and have $\ac\sim1000$-$8000\AU$ in the
other models. The travel/arrival time of the detected HVSs from the GC to its current
location is on the order of $\sim100$~Myr \citep[e.g.,][]{Brown12b}, which suggests that their companions
were captured $\sim 100$~Myr ago and the orbits of the captured companions may
have been changed due to the dynamical interactions with its environment (see Section~\ref{sec:orb_ev}). As
shown in Figure~\ref{fig:f1}, for those captured stars with $\ac\ga
10,000\AU$, the probability that they
have ejected companions with $\vinf > 700$-$1000\kms$ is negligible.

Figure~\ref{fig:f2} shows the distribution of those captured stars obtained
from each model in the $\ac$ versus $\log(1-\ec)$ plane, where $\ac$ and $\ec$ are their
semimajor axes and eccentricities achieved right after they were captured, respectively.
Relatively more captured stars with low eccentricities are produced by the
Disk-MS0 model than by the Unbd-MS0 model (see panels (a) and (b)) mainly
because the injecting binaries can be broken up at relatively larger distance
in the Disk-MS0 model due to multiple encounters. And relatively more captured
stars with low eccentricities and $\ac$ are produced in the Disk-TH model
than that in the Unbd-MS0 model (see panels (a) and (c)) because there are more
injecting binaries with mass ratio $q$ substantially less than
$1$ and $m\l \sim 7$-$15 \msun$ for a top-heavy IMF than that for the
Miller Scalo IMF. The Disk-TH2 model produces relatively more captured stars
with smaller eccentricities for any given $\ac$ than the Disk-TH0 model
(as shown in panels (c) and (d)), as those binaries are generally broken up at even
larger distances in the Disk-TH2 model because fewer binaries can approach 
the very inner region due to the steepness of the adopted $f_{\rp}(\rpi)$. However, the
eccentricities of those captured stars, even produced in the Disk-TH2 model,
are still statistically significantly higher than that of the 
observed GC S-stars.  The orbits of a number of GC S-stars, including S2, can be
directly produced in the Disk-TH0 model and the Disk-TH2 model if the injection
rate of binaries is around a few times $10^{-5}$ to $10^{-4}\pyr$ as set for those
models (see similar rates obtained by \citealt{Bromley12}).  According to Figure~\ref{fig:f2}, apparently it is extremely difficult
to produce ``GC S-stars'' with $\ec<0.8$ directly through the tidal breakup
mechanism of stellar binaries in the vicinity of the MBH. Note also that fewer
captured stars with $\ac< 1000\AU$ are produced in the Disk-TH2 model compared
with those in other models because stellar binaries are harder to approach the
innermost region than that in other models (see panel (d) in Figures \ref{fig:f1}
and \ref{fig:f2}).

\begin{figure*}
\centering
\includegraphics[scale=0.5]{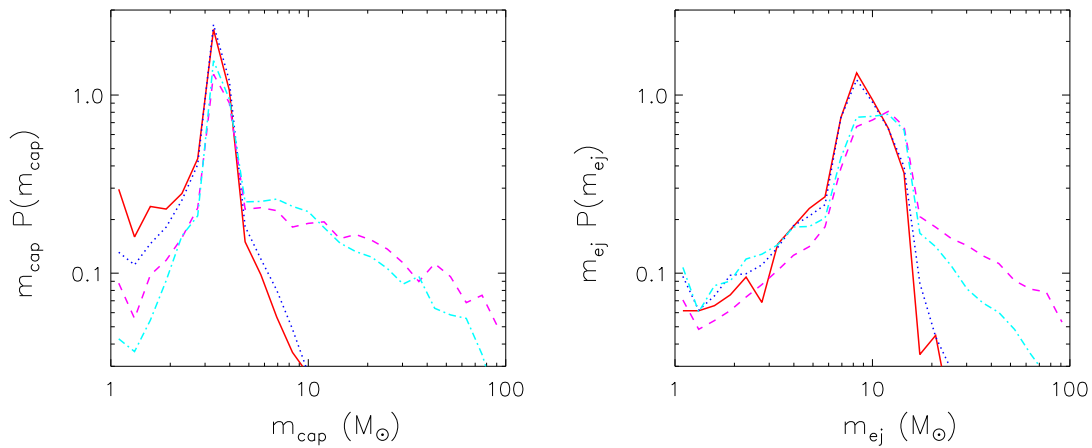} 
\caption{
Mass probability distribution $P(m_{\cap})$ of the captured companions
of the $\sim 3$-$4\msun$ HVSs (left panel)
and mass probability distribution $P(m_{\ej})$ of the ejected
companions of the $\sim 7$-$15\msun$ captured
stars (right panel). The
solid (red), dotted (blue), dashed (magenta) and dot-dashed (cyan) lines
represent the results obtained from the Unbd-MS0, the Disk-MS0,
the Disk-TH0, and the Disk-TH2 models, respectively. }
\label{fig:f3}
\end{figure*}

According to the simulations above, we find that $\sim50\%$ of the detected
HVSs should have captured companions in the GC with mass $\sim 3$-$4\msun$ as
shown in the left panel of Figure~\ref{fig:f3}. And similarly $\sim 60\%$ of the observed GC S-stars
should have ejected companions with mass $\sim 7$-$15\msun$ as shown in
the right panel of Figure~\ref{fig:f3}. To find the possible counterparts of those current
observed GC S-stars and HVSs, we will focus on the ejected stars with mass $\sim
7$-$15\msun$ in the Galactic bulge and halo and the captured stars with mass
$\sim 3$-$4\msun$ in the GC. For completeness, we also count the ejected and
captured stars with mass $\sim 4$-$7\msun$ produced in all the models (see
Table \ref{tab:t4}).

\section{Orbital evolution of the captured stars}\label{sec:orb_ev}

The orbits of the captured stars produced by the tidal breakup of stellar
binaries may evolve due to dynamical interactions with the surrounding
environments. In principle, two-body interactions between stars may cause
exchanges of their angular momenta and energy. However, the timescale of the
two-body relaxation ($\sim 10^9$~yr) is too long for it to be effective in
changing the orbits of captured stars within their main-sequence lifetime
\citep[e.g.,][]{HA06,Yuetal07}. The RR is
an important dynamical process naturally resulted from the coherent torques
between orbital averaged mass wires of stars moving in near-Keplerian potential
proposed by \citet{RT96}, which can lead to changes in both the
eccentricities (scalar RR) and orientations (vector RR) of stars moving in the
GC \citep[e.g.,][]{HA06}. The RR appears much more effective in changing the
orbital configuration of the ``GC S-stars'' than the non-resonant two-body
relaxation \citep[NR;][]{HA06,Perets09c,KT11}. Therefore, the scalar and vector
RR may be crucial in the follow-up dynamical evolution of the orbits of the
captured stars.

The vector RR timescale is $\sim 1$-$10$~Myr in the region hosting the ``GC S-stars'',
about one order of magnitude smaller than the scalar RR timescale \citep{HA06,
Yuetal07}, and thus the captured stars can evolve to an isotropic distribution
on a timescale of $\sim10$~Myr \citep{HA06, Perets09c, KT11} even if they were
originally on a plane-like structure. In this paper, we assume that the
isotropic distribution of the GC S-stars can always be reproduced through the
vector RR of the simulated captured stars within a timescale shorter than their
lifetime. It has been suggested that the high-eccentric orbits of captured
stars can dynamically evolve to that of the observed GC S-stars through
the scalar RR within $\sim20$~Myr \citep[e.g.,][]{Perets09c}. However,
previous studies assume a simple distribution of the initial eccentricities and
semimajor axes of the captured stars \citep[e.g.,][]{Perets09c, MHL10}. In this Section,
we adopt the distribution of $\ec$ and $\ac$ resulted from the injection models
studied in Section~\ref{sec:num_simu}. We follow the evolution of $\ec$ and $\ac$ by
taking both the RR and NR into account, and then compare the $\ec$ and $\ac$ distributions
of the captured stars surviving to the present time with that of the observed GC S-stars.

We adopt the ARMA model first introduced by \citet{MHL10} to perform
Monte Carlo simulations of the long-term evolution of the captured stars. In
the ARMA model, the RR phase and the NR phase are unified, and the general
relativistic (GR) precession of stars in the potential of the central MBH is also
simultaneously included. The ARMA model is characterized by the following three parameters:
(1) the autoregressive parameter $\phi_1$; (2) the moving average parameter $\theta_1$;
and (3) the parameter $\sigma_1$, which is the variance of a random variable
$\epsilon^{(1)}$ following the normal distribution. In the ARMA model,
$\epsilon^{(1)}$ represents the random walk motion of the NR phase. At a time
step of one orbital period of a star, the variation in the absolute value of
its angular momentum is 
\be
\Delta_1 J_t=\phi_1\Delta_1J_{t-1}+\theta_1\epsilon^{(1)}_{t-1}
+\epsilon^{(1)}_t,
\ee 
and
\be
\phi_1=\exp{(-\frac{\delta t_{\rm P}}{S t_\phi})},
\ee
\be
t_{\phi}=f_{\phi}{\rm min}[t_{\rm prec}(a,e),t_{\rm prec}(a,\tilde{e})],
\ee
\be
S=\frac{1}{1+\exp{\left[-k\left(e-e_{\rm crit}\right)\right]}},\footnote{
\rm There is a misprint in the original form of $S$ given by
\citet[][see their Equation 33]{MHL10}, i.e., the minus sign before $k$ was
missing. }
\ee
\be
e_{\rm crit}(a,e)=\sqrt{\frac{\ln\Lambda}{A_{\rm NR}A_{\tau}^2}}
\left(\frac{\delta t_{\rm P}}{t_{\phi}}\right),
\ee
\be
\theta_1=-\exp\left[{-\frac{f_\theta}{2}\sqrt{\frac{1}{\phi_1^2}+\phi^2_1-2+
\frac{4(1-\phi^2_1)\tau^2\delta t_{\rm P}^2}{\sigma^2_1}}}\right],
\ee
\be
\tau=A_{\tau}\frac{m_*}{\bh}\frac{\sqrt{N_{<}}}{\delta t_{\rm P}}e,
\ee
\be
\sigma_1=f_\sigma \frac{m_*}{M_\bullet}\sqrt{\frac{N_<
\ln\Lambda}{A_{\rm NR}}},
\ee
\be
f_{\sigma}=0.52+0.62e-0.36e^2+0.21e^3-0.29\sqrt{e},
\ee
where $f_{\phi}=0.105$, $f_\theta=1.2$, $k=30$, $A_{\rm NR}=0.26$, $A_{\tau}
=1.57$, $\Lambda=\bh/m_*$, $m_*=10\msun$ is the averaged mass of the field
stars, $a$ and $\delta t_{\rm P}$ are the semimajor axis and orbital period of the star,
$\tilde{e}$ is the median value of the eccentricity of the field stars and it
is $\sqrt{1/2}$ for a thermal distribution, $t_{\rm prec}$ is the combined
precession timescale for the Newtonian precession and the general relativity
precession \citep[see Equations (25), (27), and (28) in][]{MHL10}, and $N_<$ is the total
number of stars within the radius equal to the semimajor axis of the captured
star.  Adopting a simple stellar cusp model, i.e., $\rho_*\propto r^{-\alpha}$,
we have $N_<=N_{\rm h}(r/r_{\rm h})^{3-\alpha}$, where $r_{\rm h}$ represents
the radius within which the mass of stars equals the MBH mass and $N_{\rm h}
=\bh/m_*=4\times 10^5$ is the total number of field stars within $r_{\rm h}$.
Similar to \citet{MHL10}, we also assume a Bahcall--Wolf cusp ($\alpha=7/4$,
\citealt{BW76}) unless otherwise stated and correspondingly $r_{\rm h}=2.3\pc$.
The variable superscript `(1)' in the above equations means that the time
step is one orbital period of the star being investigated. In the time step of $N$ period of
the star ($N=\delta t/\delta t_{\rm P}$), the model parameters ($\phi_N$, $\theta_N$,
$\sigma_N$) can be obtained from parameters of one period ($\phi_1$,
$\theta_1$, $\sigma_1$). For further details of the ARMA model, see
\citet{MHL10}.
The ARMA model may not capture the exact dynamical physics of the system and
give the exact kinematics of each individual star; but for the purpose of our
work and the addressing problems, it should be plausible and efficient to
be applied here to obtain the evolution of the system in a statistical way.

The two-body NR is also taken into account in a way similar to that 
in \citet{MHL10}. In a time step $\delta t$, the energy change of a captured
star due to the NR is given by
\be 
\Delta E=\xi E \left(\frac{\delta t}{t_{\rm NR}}\right)^{1/2},
\ee 
where $\xi$ is an independent normal random variable with zero mean and unit
variance, and $t_{\rm NR}$ is the NR timescale in the GC given by 
\be 
t_{\rm NR}=A_{\rm NR}\left(\frac{\bh}{m_*}\right)^2\frac{1}{N_<}
\frac{1}{\ln\Lambda}\delta t_{\rm P}.
\ee 

The travel time of those HVSs discovered in the Galactic halo is $\sim
40$-$250$~Myr if HVSs were ejected from the GC \citep{Brown09a}. And
recent observations have also shown that the age of the
detected HVSs is on the order of $\sim 100$~Myr, which is consistent with
the GC origin \citep{Brown12b}. In the model of this paper, we are unifying the
formation of both the
HVSs and the GC S-stars by the tidal breakup of young stellar binaries
originated from the young stellar disk(s) in the GC. In order to be
compatible with the above observations, we assume a constant injection 
rate of stellar binaries over the past $250$~Myr\footnote{Although
the majority of the currently detected Wolf--Rayet and O/B type supergiants
and giants in the disk are young (with age of $6$~Myr or so;
\citealt{Paumard06}), the observations have also shown that there are 
many B-dwarf stars in the disk region ($0.04-0.5$~pc). For example, 
\citet{Bartko10} find $59$ B-dwarfs in the disk region and the ages of 
these dwarfs could be substantially larger than $6$~Myr. Note also that 
the observational bias on the detection of young but faint stars in the
inner parsec is largely uncertain. Thus, current observations do not exclude 
the existence of less massive stars with ages much longer than those of
the detected Wolf--Rayet and O/B type disk stars in the GC.}. 
We adopt the
numerical results of the three-body experiments for each injection model in
Section~\ref{sec:num_simu} and calculate the energy and angular momentum
evolution for each captured star by using the ARMA model. In these
calculations, we also take account of the effects of the limited lifetime of the
captured stars on the main sequence and the tidal disruption of those captured
stars moving too close to the MBH. We remove those captured stars if they move
away from the main sequence or approach the MBH within a distance of
$\rp<r^{\rm td}$, where $r^{\rm td}=(2\bh/m)^{1/3}R_{*}$ and $R_*$ is the stellar
radius. Finally, we obtain the present-day semimajor axis and eccentricity distributions of
the captured stars, which can be used to be compared to
the observational distributions and constrain the models.

\begin{figure*}
\centering
\includegraphics[scale=0.5]{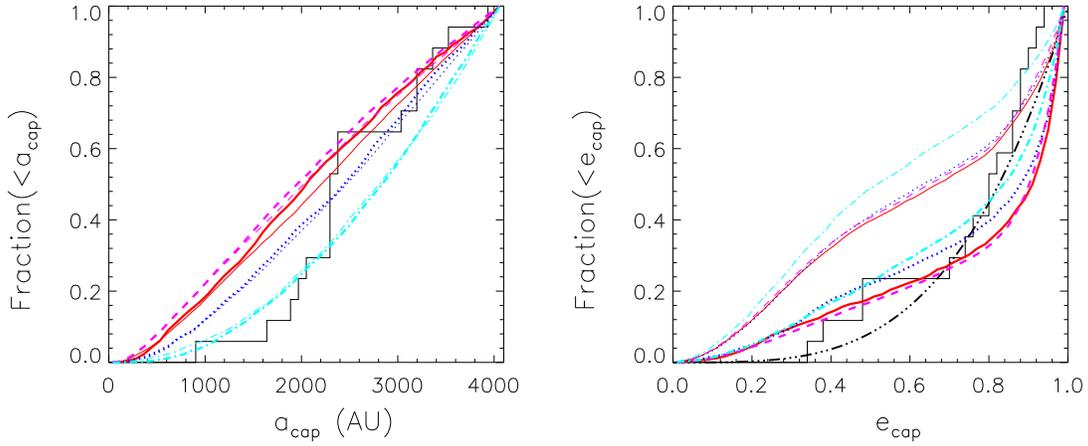}
\caption{
Cumulative distributions of the semimajor axis (left panel) and eccentricity (right panel) of the captured stars surviving to 
the present time. The histograms represent the distributions of the observed GC S-stars
\citep{Gillessenetal09}. The solid (red), dotted (blue), dashed
(magenta), and
dot-dashed (cyan) curves show the results obtained from the Unbd-MS0
model,
the Disk-MS0 model, the Disk-TH0 model, and the Disk-TH2 model, respectively. The thick and
thin curves represent the distributions of those captured
stars with mass $7$-$15\msun$ and $ 3$-$4\msun$, respectively. 
In the right panel, the dot-dot-dashed line represents the cumulative eccentricity distribution
proportional to $\ec^{3.6}$ as suggested by the observations \citep{Ghezetal08,Gillessenetal09}.
} 
\label{fig:f4}
\end{figure*}

The left panel of Figure~\ref{fig:f4} shows the cumulative distributions of $\ac$ of the captured
stars surviving to the present time and that of the observed GC S-stars. The thick lines
represent the captured stars with mass $\sim 7$-$15\msun$ surviving to the present time (the simulated GC S-stars), roughly corresponding
to the observed GC S-stars, and the thin lines represent the captured
stars with mass $\sim 3$-$4\msun$, roughly corresponding to the captured
companions of those HVSs detected in the Galactic halo. Although the lifetime
of less massive stars on the main sequence is substantially longer and thus the
dynamical evolution time is longer than that of the massive ones, the cumulative
distribution of $\ac$ of the light captured stars is only slightly different
from that of the massive ones (see the left panel of Figure \ref{fig:f4}). The slope of the $\ac$
distribution is affected most by the $f_{\rp}(\rpi)$ distribution. Relatively
fewer captured stars are produced in the inner region by the Disk-TH2 model
compared with that obtained by the other models, and the fraction of captured stars
with semimajor axes $<\ac$ is proportional to $\sim\ac^{2.0}$ for the Disk-TH2 model
but to $\sim \ac^{1.0-1.5}$ for the other three models. For those models with $\beta=0$, the $\ac$
distributions obtained from the numerical simulations is roughly consistent with the
simple estimations from Equation (\ref{eq:fac}) (e.g., the slope is $1$ for
$\alpha=-1$ and $\beta=0$ according to Equation (\ref{eq:fac})). For the Disk-TH2 model, however, the $\ac$ distribution seems flatter than the simple expectation, i.e., a slope of $3$ (for $\alpha=-1$ and $\beta=2$). The flatter slope of $f(\ac)$ resulted from the Disk-TH2 model may be
due to the effect of various mass ratios of the injecting binaries, which is
included in the numerical simulations but ignored in deriving
Equation (\ref{eq:fac}) (see panel (d) in Figure (\ref{fig:f1})). 
The Kolmogorov--Simirnov (K-S) tests find the likelihoods of
$0.02$, $0.2$, $0.05$, and $0.15$ that the $\ac$ distribution of the observed
GC S-stars is the same as that of the simulated GC S-stars for
the four models, respectively. As shown in Figure~\ref{fig:f4}, most
of the discrepancy between the observational distribution and the
simulated one is apparently near the edges of those distributions. Since
the Anderson--Darling (A-D) test may be more effective than the K-S test
and more sensitive to the distribution
edges \citep[see][]{FB12}, 
we also adopt the A-D test here and find the likelihoods are
$0.05$, $0.3$, $0.03$, and $0.32$ for the four models, respectively. 
These statistical tests suggest that the Disk-MS0 model and the
Disk-TH2 model may be more compatible with the observational $\ac$
distribution. 

For the majority of the simulated GC S-stars ($\sim 7$-$15\msun$), the relative changes
in their energy due to dynamical relaxation after their capture are less than
$\sim$30\% and their semimajor axes do not deviate much from the initial values
right after their capture. Therefore, the distribution of the semimajor axis of
currently observed GC S-stars can provide some information on their ejected
companions (see Equations (\ref{eq:dk_b}) and (\ref{eq:acap0})). For those captured
stars with mass $\sim 3$-$4\msun$, however, their relative energy changes can be
as large as $1$ mainly because of their longer lifetime and thus longer
dynamical evolution time, and Equation (\ref{eq:acap0}) is no longer reliable to
provide estimations on the velocity of the ejected companions of those less
massive captured stars by using their present-day semimajor axes.

The right panel of Figure~\ref{fig:f4} shows the cumulative eccentricity distributions obtained
from different models.  As seen from Figure~\ref{fig:f4}, the $\ec$
distributions are only slightly different for different injection models
because the initial $\ec$ are all close to $1$ in all the models. For those
captured stars with mass $\sim 3$-$4\msun$, their present eccentricities are
relatively lower than that of the observed GC S-stars because of their longer
dynamical evolution time. For those simulated GC S-stars (with mass $\sim
7$-$15\msun$) resulted from any of the four injection models, their $\ec$
distribution is similar to that of the observed GC S-stars. 
The K-S tests find a likelihood
of $\sim 0.1-0.6$ that the eccentricity distribution of the observed S-stars is
the same as that of the simulated GC S-stars for all the
four models. If alternatively adopting the A-D test, then the
likelihoods are 
$\sim 0.004, 0.02, 0.005$, and $0.13$ for the Unbd-MS0 model,
the Disk-MS0 model,
the Disk-TH0 model, and the Disk-TH2 model, respectively. According to these
calculations, the Disk-TH2 model may be more compatible with the observational $\ec$
distribution.

We note here that the simulations slightly over-produce the stars
with high eccentricities ($\ec$ close to $1$) with respect to the observations
because the RR for those captured stars with extremely high
eccentricities is quenched by the strong relativistic precession \citep{MHL10}. 
For this inconsistency, part of the reason might be the observational bias 
in detecting the GC S-stars, i.e., the stars with high eccentricities are 
less likely to be detected at $\ac\sim 0.01\pc$ \citep{Schodel03, Weinberg05, 
MHL10}; and part of the reason might be the limitation of the ARMA model. 
But the main reason of the inconsistency does not appear to be due to 
ignoration of the ``bouncing effect'' demonstrated in Figure 7 of 
\citet{Merritt11}, where the star starting from a low-eccentricity orbit 
and evolving close to a critical high-eccentricity orbit is then bounced 
back onto a low-eccentricity orbit due to the suppression of the RR by the 
fast GR precession at high-eccentricity orbits, as (1) the stars in our model 
are captured from tidal breakup of binary stars and they initially have 
eccentricities even higher than the critical eccentricities; (2) during the 
simulation period, some of the stars have evolved onto low-eccentricity orbits 
as illustrated by the distribution at the low-eccentricity end in 
Figure~\ref{fig:f4}, and the simulated stars at the
high-eccentricity end are those that evolve relatively slowly;
and (3) the suppression of the RR due to the fast GR precession has been
modeled in our work, as mentioned before (e.g., see the definition of
$t_{\rm prec}$ above).

The timescale of the RR process also depends on the mass of the
field stars \citep{MHL10,RT96}. To check this dependence, we
perform additional simulations by setting the mass of field stars to $m_*=5\msun$
or $m_*=20\msun$ but with the total mass of the field stars fixed. According to
the results of these simulations, we find that the simulated GC S-stars are on orbits with too
high eccentricities compared with the observational ones if
$m_*=5\msun$,
or on orbits with too low eccentricities if $m_*=20\msun$.

In the above calculations, a Bahcall--Wolf cusp for the background
stellar system in the GC is adopted. However, recent observations
suggested that the background stellar distribution may be core-like
rather than cusp-like (e.g., \citealt{Doetal09}). Similarly as done in
\citet{MHL10}, we also adopt $\alpha=0.5$ to mimic the effect of a
core-like distribution, and the perturbation on the GC ``S-stars'' is
assumed to be dominated by main-sequence stars (e.g., see
\citealt{Antonini12}). In such a model, we find that the resulted S-stars
on highly eccentric orbits with smaller pericenter distances are relatively
more than those obtained from the cuspy model.
The reason is that the RR is less efficient in the core-like
stellar distribution, and thus the evolution of the eccentricities of
GC ``S-stars'' is slower.

The timescale of the RR process becomes much longer at the distance of the disks.
Stars in this region are less affected by the relaxation processes and may well
preserve some of their initial orbital configurations. Observations find many B-dwarfs in
disk regions, with high eccentricities and more extended spatial distribution
than disk stars. The resulted eccentricity--distance distribution of these stars
by the RR compared with the observations may provide useful constraints on the
formation of the GC S-stars \citep{Perets10}. Our simulations also produce many
B-dwarfs in the disk region and their radial distribution is similar to the
initial input ones for the injecting binaries. However, there should also exist
B-dwarfs in the disk region that are initially formed as single stars and the
fraction of these stars is not clear yet. A detailed dynamical study for
the B-dwarfs in the disk region is complicated and beyond the scope of this
paper.

\section{Ejected HVSs in the Galactic bulge and halo}\label{sec:HVSs}

The ejected stars move away from the GC after the breakup of their progenitor
binaries and their velocities are gradually decelerated in the Galactic gravitational potential. Some of
them are unbound to the Galactic potential and can travel to the Galactic
halo and may appear as the detected HVSs if their main-sequence lifetime is
long enough compared to the travel time;
while others with lower ejecting velocity may return to the
GC. To follow the subsequent motion of the ejected components, we adopt the
Milky Way potential model given by \citet{Xue08}, which involves four
components, including the contributions from the central MBH, the Galactic
bulge, the Galactic disk, and the Galactic halo, i.e.,
\be
\Phi=\Phi_{\rm BH}+\Phi_{\rm bulge}+\Phi_{\rm disk}+ \Phi_{\rm halo},
\ee
where
\be
\Phi_{\rm BH}   =-GM_{\bullet}/r,
\ee
\be
\Phi_{\rm bulge}=-\frac{GM_{\rm bulge}}{r+r_{\rm bulge}},
\ee
\be
\Phi_{\rm disk} =-\frac{GM_{\rm disk}(1-e^{-r/b})}{r}, 
\ee
\be
\Phi_{\rm halo} =-\frac{4\pi G\rho_{\rm s} r^3_{\rm vir}}{c^3r}
\ln(1+\frac{cr}{r_{\rm vir}}),
\ee
respectively. The model parameters for the last three components are $M_{\rm
bulge}=1.5\times10^{10}\msun$, $M_{\rm disk}=5\times10^{10}\msun$, the core
radius $r_{\rm bulge}=0.6\kpc$, the scale length $b=4\kpc$, and $\rho_{\rm
s}=\frac{1}{3}\frac{c^3\rho_{\rm c}\Omega_{\rm m}\Delta_{\rm
vir}}{\ln(1+c)-c/(1+c)}$, where $\rho_{\rm c}$ is the cosmic critical density,
$\Delta_{\rm vir}=200$, $\Omega_{\rm m}$ is the cosmic fraction of matter, the
virial radius $r_{\rm vir}=267\kpc$, and the concentration $c=12$. The bulge,
the disk, and the halo potentials adopted here are all spherical. If adopting
non-spherical potentials, i.e., a triaxial bulge/halo and a flattened disk
potential, the bending effect due to the non-spherical component on the
trajectories of ejected stars is important only for those with $\vinf \la
400\kms$ on a timescale of $\ga 500$~Myr, but it is negligible for HVSs with
relatively high speeds \citep[see][]{YM07}. Note that the radial distribution of
the ejected stars surviving to the present time (and correspondingly the predicted number of the detectable
HVSs)  may be slightly different if adopting a different Galactic potential
model. 

The total number of detectable HVSs depends directly not only
on how many stellar binaries can be injected into the
immediate vicinity of the MBH, but also on the lifetime of these stars
and the detailed settings on the IMF, semimajor axis, and periapsis of the
injecting stellar binaries (see Section~\ref{sec:num_simu}). If the stellar
binaries are injected from a far away region, the injection rate can be estimated
through the loss-cone theory \citep{YT03,Perets09c}; if the injecting stellar
binaries originated from central stellar disks \citep{Luetal10,Zhang10},
the injection rate is difficult to estimate as the mechanism responsible for it
is still not clearly understood (cf.\ \citealt{MLH09}). In principle, it is
plausible to observationally calibrate the injection rate by the numbers of the
detected HVSs and GC S-stars. But this calibration becomes complicated if
considering of the uncertainties in the settings of the distributions
$f_{\ab}(\abi)$, $f_{\rp}(\rpi)$, and IMF of the injecting binaries, etc.

\begin{deluxetable*}{lccccccccc}
\tablecaption{The Number Ratio of the Simulated $3$-$4\msun$ HVSs to the Simulated GC S-stars} 
\tablehead{
\colhead{Model} &
\colhead{$\gamma$} &
\colhead{$\beta$} &
\colhead{$\frac{N_{\HVS}^{\tot}}{N_{\cap}^{\tot}}$} &
\colhead{$\frac{F^{\lt}_{\HVS}}{F^{\lt}_{\cap}}$} & 
\colhead{$F^{\rm td}_{\cap}$} & 
\colhead{$F_{\rm HVS}\obsr$} & 
\colhead{$F_{\cap}\obs$} & 
\colhead{$\frac{N_{\rm HVS}\obsr}{N_{\cap}\obs}$}} 
\startdata
  Unbd-MS0  & -2.7  &  0 & 3.0  & 8.8  & 0.59 & 0.27 & 0.51 & 27  \\
  Disk-MS0  & -2.7  &  0 & 1.8  & 8.9  & 0.51 & 0.15 & 0.41 & 12  \\
  Disk-TH0  & -0.45 &  0 & 0.22 & 10   & 0.47 & 0.19 & 0.54 & 1.4 \\
  Disk-TH2  & -0.45 &  2 & 0.27 & 9.9  & 0.34 & 0.06 & 0.21 & 1.3 \\
  Disk-IM0  & -1.6  &  0 & 0.61 & 9.5  & 0.50 & 0.16 & 0.47 & 4.0 \\
  Disk-IM2  & -1.6  &  2 & 0.75 & 9.4  & 0.22 & 0.01 & 0.03 & 2.8 
\enddata
\label{tab:t3}
\tablecomments{ 
The $N_{\rm HVS}^{\tot}$ and $N_{\cap}^{\tot}$ represent the total number
of the ejected stars with mass $3$-$4\msun$ and the total number of the captured
stars with mass $7$-$15\msun$ that are generated by the tidal breakup
of stellar binaries in the GC for each model, respectively; the $F^{\lt}_{\rm
HVS}$ and $F^{\lt}_{\cap}$ denote the fraction of the ejected and captured
stars that still remain on the main sequence of their stellar evolution at the end of our simulations;
the $F^{\rm td}_{\cap}$ denotes the fraction of the captured stars that have been tidally
disrupted by the central MBH before the end of our simulations; the $F_{\rm HVS}\obsr$
denotes the fraction of the ejected stars appear as the detected HVSs, where an
ejected star is taken as a {\it detectable} HVS if its heliocentric radial velocity in the Galactic rest frame is $|v\rf| >
275\kms$, its velocity at infinity is $\vinf >750\kms$, and its distance from the GC is in
the range of $40$-$130\kpc$; $F_{\cap}\obs$ denotes the fraction of the captured
stars that are within radii $\la 4000\AU$ from the MBH; and the $N_{\rm HVS}\obsr$ is the
total number of those {\it detectable} HVSs with mass $3$-$4\msun$ in the Galactic halo
and $N_{\cap}\obs$ is the simulated number of the captured stars with mass
$7$-$15\msun$ in the GC, which correspond to the observed ones.
}
\end{deluxetable*}

\begin{deluxetable*}{lccccccccccccccc}
\tablewidth{18.5cm}
\tabletypesize{\tiny}
\tablecaption{Predicted Numbers in Different Models}
\tablecolumns{2}
\tablehead{
\multirow{2}{*}{Model} & \multirow{2}{*}{$\gamma$} & \multirow{2}{*}{$\beta$} &
\colhead{injection rate} & 
\colhead{} & \multicolumn{3}{c}{$3$-$4\msun$} & \colhead{} &
\multicolumn{3}{c}{$4$-$7\msun$}& \colhead{} &
\multicolumn{3}{c}{$7$-$15\msun$}  \\
\cline{6-8} \cline{10-12} \cline{14-16}
   &   & & $(10^{-5}\,{\rm yr^{-1}})$& & $N_{\rm HVS}\obsr$ & $N_{\rm HVS}\obsp$  & $N_{\cap}\obs$ & & 
             $N_{\rm HVS}\obsr$ & $N_{\rm HVS}\obsp$  & $N_{\cap}\obs$ & & 
             $N_{\rm HVS}\obsr$ & $N_{\rm HVS}\obsp$  & $N_{\cap}\obs$
}
\startdata
Unbd-MS0   &-2.7  & 0 & 13  (2.2) & &504 (79)&189 (30)& 86 (13)& &178 (28)& 58  (9) &33 (5)  & & 60 (9) & 33 (5) & 17  (3)\\
Disk-MS0   &-2.7  & 0 & 7.0 (2.7) & &177 (79)& 66 (29)& 82 (36)& & 62 (28)& 21  (9) & 32 (14)& & 29 (13)& 17 (8) & 17  (8)\\
Disk-TH0   &-0.45 & 0 & 1.8 (5.8) & &23  (79)&8   (27)& 11 (37)& & 28 (97)& 8  (30) & 11 (39)& &29 (103)& 19 (66)& 17 (60)\\
Disk-TH2   &-0.45 & 2 & 6.2  (23) & &18  (79)&7   (31)& 8  (34)& & 16 (71)& 5  (23) & 9  (40)& & 20 (87)& 15 (64)& 17 (75)\\
Disk-IM0   &-1.6  & 0 & 3.1 (3.6) & &61  (79)& 22 (29)& 28 (36)& & 37 (48)& 12 (15) & 18 (23)& & 29 (38)& 18 (23)& 17 (22)\\
Disk-IM2   &-1.6  & 2 & 60  (100) & &48  (79)& 21 (35)& 32 (52)& & 24 (39)& 7  (11) & 18 (29)& & 20 (33)& 14 (23)& 17 (28)
\enddata
\label{tab:t4}
\tabletypesize{\footnotesize}
\tablecomments{
The numbers of the simulated {\it detectable} HVSs and the captured stars surviving in the
GC at the present time in different mass ranges, obtained from different models.
The injection rate of stellar binaries is assumed to be a constant over the
past $250$~Myr, which enables the production of $17$ simulated GC S-stars (or $79$
unbound $3\sim4\msun$ HVSs (numbers in the brackets)). The $N_{\rm HVS}\obsr$ denotes
the number of HVSs with $|v\rf|>275\kms$ and $\vinf>750\kms$; and the $N_{\rm
HVS}\obsp$ denotes the number of the HVSs with proper motion $\ge 5\maspyr$ in the heliocentric rest frame and
$\vinf>750\kms$. In order to compare to the observations, a simulated HVS
with mass of $3$-$4\msun$ is counted as a {\it detectable} HVS additionally if its
distance to the GC is in the range of $40-130\kpc$. The $N_{\cap}\obs$ is the
total number of the captured stars surviving in the GC with present-day $\ac\la 4000\AU$ in the
simulations.
}
\end{deluxetable*}

\subsection{The Numbers of the HVSs/GC S-stars and Their Number Ratio}\label{subsec:number}

The $3$-$4\msun$ HVSs detected in the Galactic halo and the GC S-stars should be
linked to each other under the working hypothesis of this paper. The total
numbers of the simulated $3$-$4\msun$ HVSs and $7$-$15\msun$ GC S-stars depend not only on the injection rate
of stellar binaries but also on the detailed settings on the injection models.
However, the number ratio of the simulated $3$-$4\msun$ HVSs to the
simulated $7$-$15\msun$ GC S-stars may depend
only on the details of the injection models described in Section~\ref{sec:num_simu},
but not on the injection rate of stellar binaries. Any viable model should produce
a number ratio compatible with the observations on the HVSs and the GC S-stars,
and thus this number ratio may provide important constraints on the models. 

We obtain both the numbers of the simulated HVSs and the captured stars surviving to
the present time and their number
ratio by Monte Carlo simulations as follows. First, we obtain the total
number of initially captured (or ejected) stars, $N_{\cap}^{\tot}$ (or
$N_{\HVS}^{\tot}$), with mass in the ranges of $3$-$4\msun$, $4$-$7\msun$, and
$7$-$15\msun$, respectively. To do this, we assume a constant injection rate of
binaries and randomly set the injection events over the past $250$~Myr, and for
each injection event we randomly assign it to a three-body experiment conducted
in Section~\ref{sec:num_simu} and adopt the results from the experiment.
Second, we consider the limited lifetime of each ejected and captured star, the
motion of each ejected star in the Galactic potential, and the dynamical
evolution of each captured star, and then obtain the fraction of the ejected
stars ($F_{\rm HVS}^{\lt}$) that still remain on the main sequence at the present 
time or the similar fraction for the captured stars ($F_{\cap}^{\lt}$),
and the fraction of the
captured stars ($F_{\cap}^{\rm td}$) that have already been tidally disrupted
until the present time. Third, we consider the kinematic selection criteria and obtain the
fractions of those ejected and captured stars according to given observational selection criteria,
i.e., $F_{\rm HVS}\obs$ and $F_{\cap}\obs$, respectively. The selection criteria
are similar to those adopted in selecting the detected HVSs and GC S-stars, i.e.,
an ejected star is labeled as a {\it detectable} HVS if its heliocentric radial velocity in the Galactic rest frame is $|v\rf|
>275\kms$ or its proper motion in the heliocentric rest frame is $\ge 5\maspyr$, and its velocity at infinity is
$\vinf >750\kms$, and captured stars with semimajor axis $\la 4000\AU$
are counted as {\it detectable} GC S-stars. For those simulated HVSs with mass $
3$-$4\msun$, we put an additional cut on their distances from the GC, i.e., from
$40$ to $130\kpc$, in order to compare them to current observations.
Finally, we obtain the number ratio of the {\it detectable} HVSs to the
{\it detectable} GC S-stars for each model as 
\be 
\frac{N_{\rm HVS}\obs}{N_{\cap}^{\rm
obs}}= \frac{N_{\HVS}^{\tot}}{N_{\cap}^{\tot}}\times \frac{F^{\rm
lt}_{\HVS}}{F^{\lt}_{\cap}}\times \frac{F_{\rm HVS}\obs}{(1-F^{\rm
td}_{\cap}){F_{\cap}\obs}}.  
\ee 
The simulation results are listed in
Tables~\ref{tab:t3} and~\ref{tab:t4} for each model. 

Here we comment on a few factors that affect the predicted numbers of the
simulated {\it detectable} HVSs and GC S-stars and consequently the number ratio of
these two populations. (1) The difference between the lifetime of the
simulated HVSs and GC S-stars: the detected HVSs are in the mass range of $\sim
3$-$4\msun$, which are substantially smaller than that of the observed GC S-stars ($\sim
7$-$15\msun$). The lifetime difference leads to an enhancement in the
number ratio of the {\it detectable} HVSs to the GC S-stars roughly by a factor of $10$
under the assumption of a constant rate of injecting stellar binaries into the
vicinity of the MBH over the past $250$~Myr. (2) The place where the
stellar binaries are originated: in the Unbd-MS0 model, relatively more
stars with high velocities (e.g., $>1000\kms$) are generated than those in the
other models.  (3) The distribution of pericenter distance of those
injecting stellar binaries, which is related to the speed of the migration or
diffusion process of those binaries to the immediate vicinity of the central
MBH: a change in this distribution may result in either a significant increase
or decrease in both the number of HVSs and that of captured stars, but their
number ratio is not affected much.

The complete survey conducted by \citet{Brown09a} has detected $14$
unbound HVSs with mass $\sim 3$-$4\msun$ in a sky area of $\sim 7300$~deg$^2$,\footnote{
One sdO type star with mass $\sim 1\msun$ and another massive HVS in the
southern hemisphere with mass $\sim 9\msun$ in the survey are not included in
the number.} 
and the total number of similar HVSs in the whole sky should be $\sim
79\pm21$.\footnote{Considering of the new results on searching HVSs reported by \citet{Brown12},
this number could be slightly higher, i.e., $95\pm 23$. And if assuming that the detected HVSs were originated from two
disk-like stellar structures, i.e., the CWS disk plane and the Narm plane, as
suggested by \citet{Luetal10}, the expected total number of HVSs with mass
$\sim 3$-$4\msun$ is $75\pm28$. The number ratios of the detected HVSs to the
GC S-stars become $\sim 2.2$-$5.0$.} Observations have revealed $17$ GC S-stars within a
distance of $\sim 4000\AU$ from the central MBH. The number ratio of the
detected HVSs to the GC S-stars is $\sim3.4$-$5.9$.  The number ratio resulted from
any of the top four injection models listed in Table~\ref{tab:t2} is inconsistent
with the observational ones.  Both the Unbd-MS0 model and the
Disk-MS0 model give a number ratio substantially larger than that inferred
from observations, while the Disk-TH0 model and the Disk-TH2 model give
too small number ratios. 

\citet{Figer99} suggest that the IMF of young star clusters in the GC, i.e.,
the Arches cluster, may be top-heavy and the IMF slope is $\gamma\sim -1.6$,
although some later studies argued that the IMF of Arches cluster may be still
consistent with the Salpeter one by assuming continuous star formation
\citep[e.g.,][]{Lockmann10}. Considering of this, we adopt an IMF with a slope
of $\gamma=-1.6$ and perform two additional injecting models, i.e., ``Disk-IM0'' and
``Disk-IM2'', as listed in Tables~\ref{tab:t3} and \ref{tab:t4}. Our
calculations show that the number ratios produced by the two models
are close to the observational ones. Obviously, the number ratio
of the simulated HVSs to GC S-stars is significantly affected by the adopted IMF.
Adopting a steeper IMF may lead to a larger number ratio.

The velocity distribution of the detected HVSs suggests a slow
migration/diffusion of stellar binaries into the immediate vicinity of the
central MBH \citep[see detailed discussions in][]{Zhang10}. The Disk-IM2 model
can produce a velocity distribution similar to the observational ones, and the
large $\beta$ adopted in this model also suggests a slow migration/diffusion of
binaries into the vicinity of the central MBH. However, all the other models
appear to generate too many HVSs at the high-velocity end and thus a too flat velocity
distribution compared with the observational ones.  This inconsistency could
be due to many factors. For example, (1) the velocity distribution of the
detected HVSs could be biased due to either the small number statistics and/or
the uncertainties in estimating the observational selection effects; (2) the
uncertainties in the initial settings of the injecting binaries could also lead
to some change in the velocity distribution, e.g., the velocity distribution
may be steeper if the semimajor axis distribution of stellar binaries is
log-normal, rather than follow the \"{O}pik law \citep{Sesana07,Zhang10}; and (3) the
injection of binaries to the vicinity of the central MBH may be quite different
from the simple models adopted in this paper because of the complicated
environment of the very central region, e.g., the possible existence of a
number of stellar-mass BHs or an intermediate-mass black holes (BHs) within $100\AU$.

For each model, the injection rate of stellar binaries can be calibrated to
produce the numbers of the observed HVSs and GC S-stars.
Under the initial settings for each model described in Section~\ref{sec:num_simu}, the
numbers of the simulated HVSs and the GC S-stars, similar to the detected ones,
are listed in Table~\ref{tab:t4}.  The calibrated injection rates are also
listed in Table~\ref{tab:t4} for each model and they are roughly on the order
of $\sim10^{-3}$ to $10^{-5}~\pyr$. However, one should be cautious about that this
injection rate depends on the initial settings on the distributions
of the semimajor axes and the periapses of the injecting binaries. According 
to our simulations, many of the injected binaries are not tidally broken up, and the
breakup rate is a factor of $3$-$6$ times smaller than the injection rate
for those models studied in this paper, i.e., about a few times
$10^{-4}$ to $10^{-5}~\pyr$, which is consistent with the estimates by \citet{Bromley12}.

\subsection{The Ejected Companions of the GC S-stars}

\begin{figure*}
\centering
\includegraphics[scale=0.5]{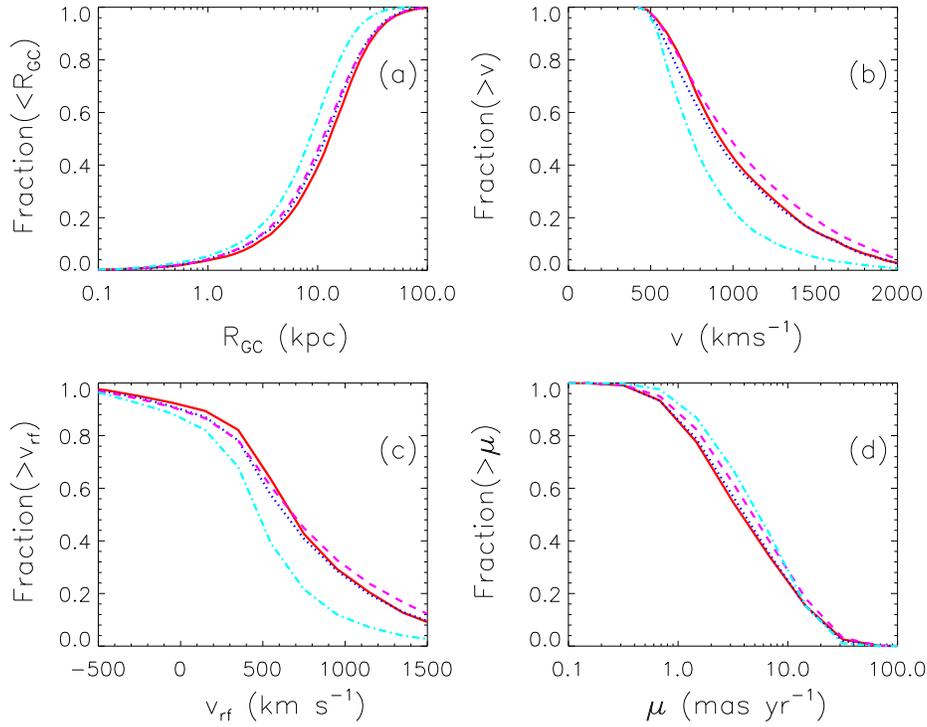}
\caption{
Cumulative distributions of the Galactocentric distance $R_{\rm GC}$
(panel (a)), the velocity in the
Galactocentric rest frame (panel (b)), the heliocentric radial
velocity in the Galactic rest frame (panel (c)), and the proper motion in
the heliocentric rest frame (panel (d)), of
those simulated unbound HVSs ($v_\infty>750\kms$) that were initially 
associated with the ``GC S-stars''. The solid (red), dotted (blue), dashed (magenta),
and dot-dashed (cyan) curves are for the Unbd-MS0 model, the Disk-MS0
model,
the Disk-TH0 model, and the Disk-TH2 model, respectively.
} 
\label{fig:f5}
\end{figure*}

\begin{figure}
\centering
\includegraphics[scale=0.5]{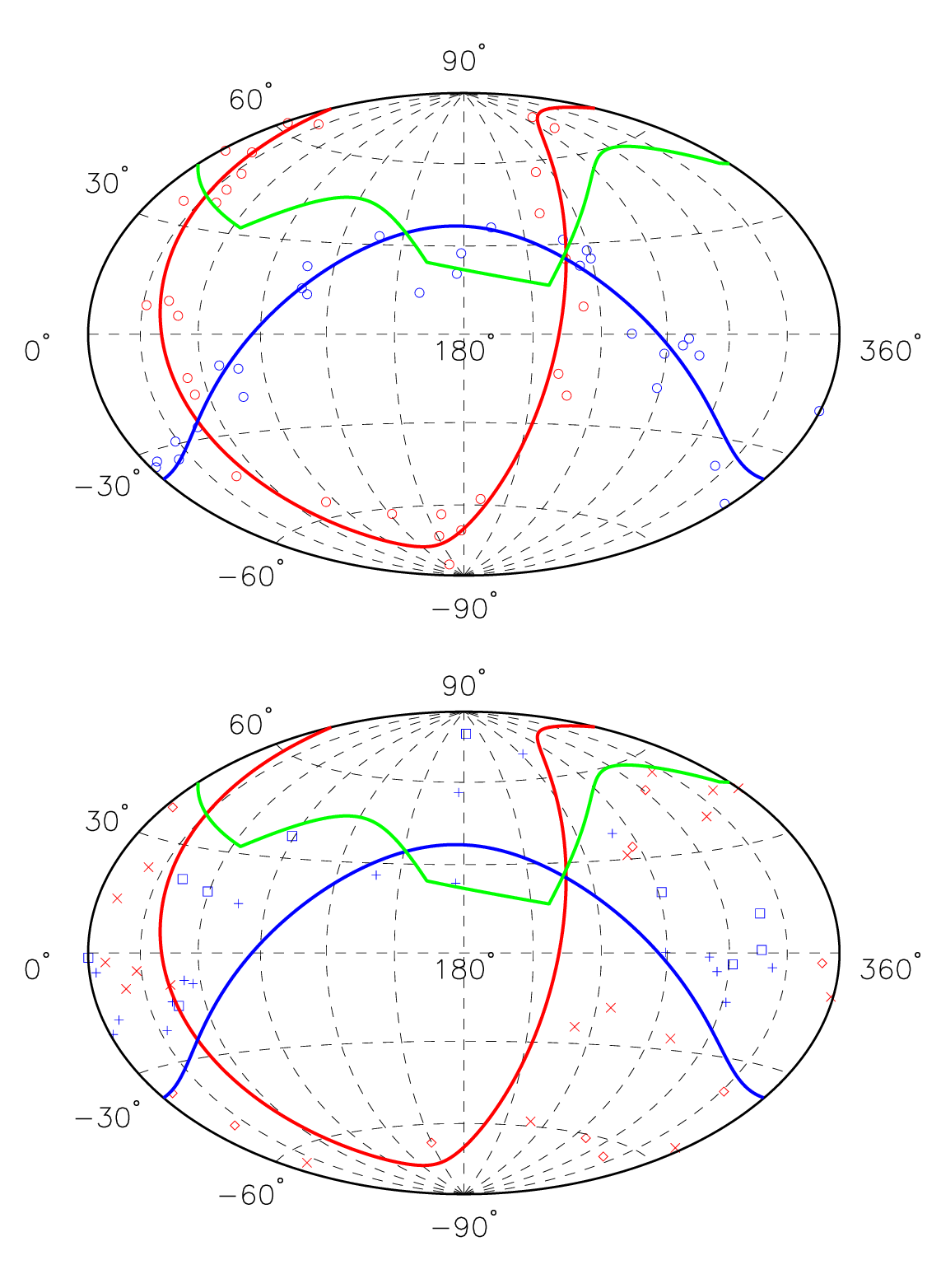}
\caption{
Spatial distribution of the simulated unbound HVSs with mass $7$-$15\msun$ and
$v_\infty>750\kms$ for the Disk-IM2 model. The distribution is expressed by a
Hammer-Aitoff projection in the Galactic coordinates. 
In the top panel, the positions of the HVSs are projected to infinity
from the GC. The solid red and blue curves represent the CWS and NARM
disk planes, respectively, which are also projected to infinity. The region above the green curve shows the area
surveyed by \citet{Brown09a} in the northern hemisphere. The open red and blue
circles represent the unbound HVSs originated from binary stars on the CWS and NARM disk
planes, respectively. For illustration purpose, all the $34$ simulated  HVSs are
shown in the top panel (see Table \ref{tab:t4}).  The injection rates are assumed
to be the same for the injections from the CWS and the NARM planes and thus the number of HVSs
associated with the two planes are also the same. In the bottom panel, the
positions of the HVSs are not projected to infinity.
The solid red curves and the solid green curve are the same as those for
the top panel. The red crosses and blue plus symbols represent those unbound
HVSs with $|v\rf|\ge 275\kms$ injected from the CWS and NARM disk planes, respectively.
The red open diamonds and the blue open squares represent the unbound HVSs with
proper motion $\mu>5\maspyr$ injected from the CWS and the NARM disk planes, respectively.
}
\label{fig:f6}
\end{figure}

The HVSs discovered in the Galactic halo are typically in the mass range $\sim
3$-$4\msun$. HVSs with other masses should also be populated in the Galactic bulge
and halo. In this Section, we investigate the properties of the simulated HVSs
with mass $\sim 7$-$15\msun$, which are most likely to be the companions of the
``GC S-stars''.

Panel (a) of Figure~\ref{fig:f5} shows the cumulative distribution of the Galactocentric
distances of the high-mass ($\sim 7$-$15\msun$) HVSs. The predicted numbers of
these HVSs from different models are listed in Table~\ref{tab:t4}. The total
number of the simulated unbound HVSs, as companions of the population of the GC S-stars,
is $\sim 20$-$30$ according to the Disk-IM2
model, which is able to reproduce the numbers of the observed HVSs
and GC S-stars. The majority of these unbound HVSs are at distances of a few
to a few tens $\kpc$ from the GC, which are much closer to the GC than
the detected $3$-$4\msun$ HVSs mainly because a high-mass
star has a shorter main-sequence lifetime and thus the distance it can travel
within the lifetime is small. The close distances of these HVSs
from the GC suggest that the velocity vectors of many of them are not along
our line of sight. Their three-dimensional velocities range from $500\kms $
to $2000\kms$ in the Galactocentric rest frame (see panel (b) in
Figure \ref{fig:f5}); and
their heliocentric radial velocities in the Galactic rest frame range from $-500\kms$
to $1500\kms$ (see panel (c) in Figure \ref{fig:f5}), where the negative and positive
velocities represent moving toward and away from the Sun, respectively.
Compared with the ejection velocities of the $3$-$4\msun$ HVSs, those
of the $7$-$15\msun$
HVSs are relatively higher because of their higher mass (see
Equation (\ref{eq:vhvs})).  Most of the HVSs have proper motions in the heliocentric rest
frame as large as $\maspyr$
to a few tens $\maspyr$ (see panel (d) in Figure~\ref{fig:f5}). These HVSs are
bright enough to be detected at a distance less than a few tens $\kpc$ by
future telescopes, such as, the {\it Global Astrometric Interferometer for
Astrophysics} spacecraft ({\it Gaia}), and their proper motions are also large enough
to be measured.

Figure~\ref{fig:f6} shows the sky distribution of the simulated HVSs
with mass$\sim 7$-$15\msun$,
which may represent the ejected companions of the ``GC S-stars'', in the Galactic coordinates by a Hammer Aitoff projection.
In the top panel of Figure~\ref{fig:f6}, the positions of the HVSs are projected to infinity from the
GC, which are consistent with being located close to the
CWS disk plane and the Narm plane (also projected to infinity) as expected.
In the bottom panel of Figure~\ref{fig:f6}, the positions of the HVSs are not projected to infinity.
As seen from the bottom panel, the simulated HVSs with mass $\sim 7$-$15\msun$ lie in the area below the projected
curves of the CWS disk plane and the Narm plane, and most of these
high-mass HVSs reside out of the area surveyed by \citet[][]{Brown09b}. Our calculations show that less than $10\%$ of the simulated
unbound HVSs with mass $7\sim15\msun$ are located in the survey area. This may
be the reason that none of those high-mass HVSs, possibly the companions of the 
``GC S-stars'', has been discovered in the survey area.  If the HVSs are initially originated
from the CWS disk and the Narm plane, the HVSs with high radial velocities are
also relatively rare in the direction close to the disk normals, i.e., $(l,b)
= (311\arcdeg, -14\arcdeg)$ and ($176\arcdeg, -53\arcdeg$) while the HVSs with
high proper motions ($\sim 20\maspyr$) is relatively numerous in that direction
because the velocity vectors of HVSs are close to be perpendicular to the disk
normals. Surveys of HVSs in the southern sky with SkyMapper and others may find
such massive HVSs, as the possible companions of the ``GC S-stars'', and provide crucial 
evidence for whether those GC S-stars are produced by the tidal breakup of stellar binaries.

Note that the spatial distribution of the HVSs discussed in this section is 
directly related to the assumption that the injecting stellar binaries are
originated from two disk-like stellar structures similar to the CWS disk in
the GC. However, the other properties of the HVSs or the GC S-stars discussed
in this paper are affected by whether the injecting stellar binaries 
are bound to the central MBH or not, but not affected by whether they are 
initially on the CWS disk plane or not.

\section{The innermost captured star}\label{sec:innermost}
 
The Unbd-MS0, Disk-MS0,
Disk-TH0, Disk-TH2, Disk-IM0, and Disk-IM2 models roughly produce $147$, $111$,
$21$, $18$, $52$, and $46$ captured stars surviving to the present
time and with mass in the range of $3$-$7\msun$,
less massive than that of the GC S-stars, within a distance of $\sim 4000\AU$ from
the MBH (see Table \ref{tab:t4} and Section~\ref{sec:HVSs}). The above numbers
are obtained by calibrating the injection rate of the stellar binaries over the past
$250$~Myr to generate $17$ simulated GC S-stars similar to the observational number.
The captured stars with mass in the range of $ 3$-$7\msun$
could be detected by the next generation telescopes, e.g., the Thirty Meter
Telescope (TMT) or the European Extremely Large Telescope (E-ELT). These 
low-mass stars are potentially important probes for testing the GR
effects near an MBH, if they are closer to the central MBH than S2.
In this section, we estimate the probability distribution of the innermost
captured low-mass stars ($\sim 3$-$7\msun$) by Monte Carlo realizations based on the
calibrated injection rate. 

\begin{figure*}
\centering
\includegraphics[scale=0.5]{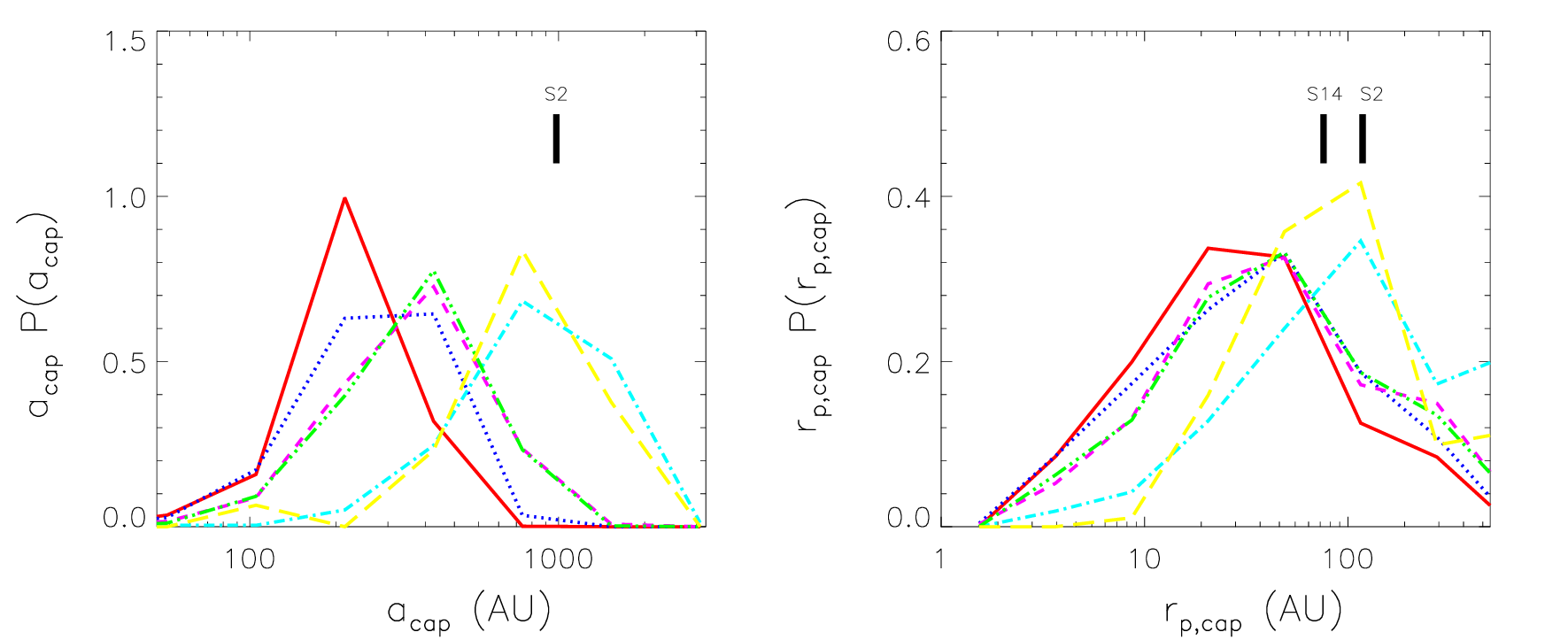}
\caption{
Probability distributions of the semimajor axis (left panel, $P(a_{\cap})$) and the pericenter distance (right panel, $P(r_{\rm p,cap})$) of the innermost captured star
with mass in the range $\sim 3$-$7\msun$ (lower than the masses of the GC S-stars) at the present time. The solid
(red), dotted (blue), short-dashed (magenta), dot-dashed (cyan),
triple -dot-dashed
(green), and long-dashed (yellow) lines show the results obtained from the
Unbd-MS0, Disk-MS0, Disk-TH0, Disk-TH2, Disk-IM0, and Disk-IM2 models,
respectively. Note here that these probability functions rely on the
estimation of the numbers of the simulated detectable captured stars (see
Section~\ref{sec:HVSs}). The estimates are obtained by calibrating the
injection rate of stellar binaries over the past $250$~Myr to generate the same number (17)
of simulated GC S-stars surviving to the present time as that of the observed ones. 
For reference, the position of S2 (or S14) is labeled in the figure.
The probabilities that
the innermost star is located within the S2 orbit (i.e., $\ac<1000$AU) are
$0.99$, $0,99$, $0.99$, $0.61$, $0.99$, and $0.73$ for the six models, respectively. 
And the probabilities that the pericenter distance of the inner most star is smaller than
that of S2 (S14),  i.e., 120\,AU (76\,AU), are $0.86$ ($0.83$), $0.81$ ($0.75$), 
$0.78$ ($0.70$), $0.55$ ($0.38$), $0.80$ ($0.70$), and $0.62$ ($0.46$) for the six models, respectively.
}
\label{fig:f7}
\end{figure*}

The left panel of Figure~\ref{fig:f7} shows the probability distributions of
the semimajor axis of the innermost captured star with mass $\sim 3$-$7\msun$
resulted from different injection models. For those models
adopting $\beta=0$, the resulted innermost captured star is typically on an 
orbit with semimajor axis $\sim 300 \AU$, and the probability that its 
semimajor axis is less than that of S2 is $\sim 99\%$. For the other models 
adopting $\beta=2$, the resulted innermost captured star is on an orbit with 
semimajor axis of $\sim 300$-$1500 \AU$ and the probability that its semimajor 
axis is smaller than that of S2 is $\sim 60\%$-$70\%$. The probability to capture 
a star within the orbit of S2 is larger for the $\beta=0$ models than
for the $\beta=2$ models. The reason is that relatively more 
stellar binaries can be injected into the immediate vicinity of the central 
MBH and thus more stars can be captured onto orbits with smaller semimajor axes
(see Table \ref{tab:t4}) in the models adopting $\beta=0$ than that adopting
$\beta=2$. We conclude that the probability of a less massive star
($3$-$7\msun$)
existing within the S2 orbit is at 
least $61\%$ and can be up to $99\%$, which may be revealed by future observations and then offer important tests 
to general relativity. 

The right panel of Figure~\ref{fig:f7} shows the probability distribution
of the pericenter distance of the innermost captured low-mass star ($\sim
3$-$7\msun$) resulted from different injection models. For the
injection models adopting $\beta=0$, the pericenter distance distribution
is concentrated within $50\AU$; while for the other models adopting
$\beta=2$, the expected pericenter distance is broadly distributed
over 10--200\,AU.  Nevertheless, the probability that the pericenter
distance of the innermost captured star with mass $\sim 3$-$7\msun$ is less
than that of S2 (and S14) is still significant, i.e., $\ga 55\%$ (or $38\%$). The innermost captured 
star may have its semimajor axis and pericenter distance both significantly smaller 
than those of S2, therefore, the GR effects on its orbit may be much more 
significant than that on S2. 

If taking into account the captured stars with even lower masses, e.g., $1\msun$, the
number of the expected captured stars surviving to the present time becomes much
larger, especially for those models with large $\gamma$. For example, the
numbers of the captured stars with mass 1--7$\msun$ are $907$, $841$, $49$, $39$,
$214$, and $117$ for the six models, respectively. For those lower mass captured
stars, the semimajor axis and the pericenter of the innermost one could be
even closer to the central MBH.

Some stars may be transported to the vicinity of the central MBH by some
mechanisms other than the tidal breakup of stellar binaries. It is possible 
that some of these stars, with their origins different from the captured stars
discussed above, exist within the S2 orbit, but which is beyond the scope of the
study in this paper.

\section{Conclusions}\label{sec:conclusion}

In this paper, we investigate the link between the GC S-stars and the
HVSs discovered in the Galactic halo under the hypothesis that they are both
the products of the tidal breakup processes of stellar binaries in the vicinity
of the central MBH. We perform a large number of the three-body experiments and
the Monte Carlo simulations to realize the tidal breakup processes of stellar
binaries by assuming a continuous binary injection rate over the past
$250$~Myr, and adopting several sets of initial settings on the injection of
binaries. After the tidal breakup of a binary, we follow the dynamical
evolution of the captured components in the GC by using the ARMA model (see
\citealt{MHL10}), which takes into account both the RR and NR processes, and we also
trace the kinematic motion of the ejected component in the Galactic
gravitational potential. 

The properties of the ejected and captured components of the tidally broken-up
binaries are naturally linked to each other as they are both the products of
tidal breakup of binaries. For those HVSs discovered in the Galactic halo with
mass $\sim 3$-$4\msun$ and $\vinf \sim 700$-$1000\kms$, their companions are
expected to be captured onto orbits with semimajor axis in the range $\sim
1000$-$8000\AU$; for the observed GC S-stars with semimajor axis $\sim
1000$-4000$\AU$ in the GC, their companions are expected to be ejected out to
the Galactic bulge and halo with $\vinf \sim 500$-$2000\kms$. 

The energy of the captured stars evolves with time because of their dynamical
interactions with the environment. For the captured stars with mass
$\sim 7$-$15\msun$, the differences between their present-day energy and their initial
ones are no more than $30\%$; for the captured stars with mass $\sim
3$-$4\msun$,
however, the difference can be by order of unity.  Therefore, the current
semimajor axis distribution of the GC S-stars may provide a good estimation on the
velocity distribution of their ejected companions (e.g., Equation
(\ref{eq:acap0})),
but that of the captured stars with mass $\sim 3$-$4\msun$ does not. The
eccentricities of the ``GC S-stars'' ($\sim 7$-$15\msun$) are close to $1$
right after the capture and may evolve to low values, and the eccentricity
distribution of these simulated GC S-stars at the present time could be
statistically compatible with the observational ones of the GC S-stars attributed
to the RR processes. For those captured stars with mass $\sim 3$-$4 \msun$, their
eccentricities can evolve to even lower values at the present time compared with the high-mass
GC S-stars ($\sim 7$-$15\msun$) because they interact with the environment for
a longer time.

To reproduce both the numbers of the detected HVSs and GC S-stars, the injection
rate of binaries need to be on the order of $10^{-4}$ to $10^{-5}\pyr$ and the IMF of the
primary components is required to be somewhat top-heavy with a slope of $\sim
1.6$. For the injection models that can reproduce the observational results on both
the GC S-stars and the HVSs, including the distributions of the semimajor axes and
eccentricities of the GC S-stars, the spatial and velocity distributions of the
detected HVSs, and the number ratio of the HVSs to the GC S-stars, the expected
number of the $\sim 3$-$7\msun$ captured companions is 
$\sim 50$ within a distance of $\sim 4000\AU$ from the central MBH. Future 
observations on the low-mass captured stars may provide a crucial check on whether 
the GC S-stars are originated from the tidal breakup of stellar binaries. 

The companions of the HVSs, which are captured by
the central MBH, are usually less massive than that of the GC S-stars
($\sim 7$-$15\msun$). The semimajor axis of the innermost captured star with mass $\sim
3$-$7\msun$ is $\sim 300$-$1500\AU$, and the probability that it is smaller than that of
S2 is $\sim 70\%$-$90\%$ for the $\beta=2$ models and $\sim 99\%$ for the $\beta=0$ models. 
The pericenter distance of the innermost captured star with mass $\sim
3$-$7\msun$
is $\sim 10$-$200\AU$ and the probability that it is smaller than that of S2 (or S14)
is also significant, i.e., $\ga 55\%$ (or $38\%$).
The existence of such a star will provide a probe for testing the
GR effects in the vicinity of an MBH. Future observations by the next
generation telescopes, such as, TMT or E-ELT, will be able to investigate the
existence of such a star, and provide important constraints on the nature
of the central MBH if such a star is detected.

The number of the ejected unbound companions of the ``GC S-stars'' (see the definition 
of the ``GC S-stars'' at the end of Section~\ref{sec:overview}) is roughly
$\sim 20$-$40$ 
and the majority of these ejected stars are located within a
distance of $\sim 20$~kpc from the GC. The number of these ejected companions is substantially
larger than the number of observed GC S-stars mainly because the observed GC S-stars are only a fraction of
the ``GC S-stars'' and the rest of the ``GC S-stars''
were tidally disrupted and do not survive today (see Table~\ref{tab:t4}).
Their heliocentric radial velocities in the Galactic rest frame
range from $\sim -500\kms$ to $\sim 1500\kms$ and their
proper motions in the heliocentric rest frame can be as large as $\sim20\maspyr$. These high-mass ejected stars are bright enough
to be detected at a distance less than a few ten kpc and their proper motions
are also large enough to be measured by future telescopes, such as
{\it Gaia}.  The
majority of the ejected companions of the GC S-stars lie outside the area surveyed
by \citet{Brown09a} for our observers located at the Sun.  

\acknowledgements
We thank the referee and the scientific editor, Eric Feigelson, for
helpful comments and suggestions. We thank Warren Brown for useful
comments on the paper and are grateful to Ann-Marie Madigan for
helpful communications on the ARMA model for the dynamical evolution
of the captured stars. This work was supported in part by the National
Natural Science Foundation of China under nos. 10973001 and 10973017,
and the BaiRen program from the National Astronomical Observatories,
Chinese Academy of Sciences.

{\bf \it Note added in proof.} After the submission of this paper, the
following two new observational results have been reported, which are
relevant to this work.  (1) Meyer et al. (2012, Sci., 338, 84)
discovered a faint star, S0-102, which is orbiting the MBH in the GC
with shortest-known-period (11.5 yr).  The existence of such a star is
consistent with our predictions shown in Figure 7. (2) Lu et al.
(2013, ApJ, 764, 155) estimate the initial mass function for stellar
populations in the central 0.5 pc of the Galaxy and find it is
top-heavy with a slope of $-1.7\pm 0.2$, which is consistent with the
requirement by our model to re-produce the number ratio of HVSs to GC
S-stars (see Table 4 and Section 5.1).

\end{document}